\PassOptionsToPackage{colorlinks=true
	,urlcolor=blue
	,anchorcolor=blue
	,citecolor=blue
	,filecolor=blue
	,linkcolor=blue
	,menucolor=blue
	,linktocpage=true
	,pdfproducer=medialab
	,pdfa=true}{hyperref}
\documentclass{article}
\pdfoutput=1
\usepackage{jcappub}
\usepackage{graphicx}
\usepackage[latin1]{inputenc}
\usepackage{amsmath,upgreek}
\usepackage{amssymb}
\usepackage{a4wide}
\usepackage{url,hyperref}
\newcommand{\pd}[3]{\frac{\partial^{#3} #1}{\partial {#2}^{#3}}} 
\newcommand{\td}[3]{\frac{d^{#3} #1}{d {#2}^{#3}}} 
\renewcommand{\v}[1]{\ensuremath{\mathbf{#1}}} 
\newcommand{\gv}[1]{\ensuremath{\mbox{\boldmath$ #1 $}}}
\renewcommand{\bar}[1]{\ensuremath{\overline{#1}}}

\title{An excess of excesses examined via dark matter radio emissions from galaxies}
\author{Geoff Beck}

\affiliation{School of Physics, University of the Witwatersrand, Private Bag 3, WITS-2050, Johannesburg, South Africa}

\emailAdd{geoffrey.beck@wits.ac.za}

\abstract{Cosmic-ray and gamma-ray observations have yielded several notable excesses that often lend themselves to explanation by various dark matter annihilation/decay models. In particular,  the AMS-02 anti-proton and positron excesses have continued to grow more robust with the collection of more data. This is supplemented by gamma-ray excesses in the Galactic Centre and a high-energy break in the spectrum of electron/positron cosmic rays seen by DAMPE. In this work we carefully model the magnetic field environments of M31 and M33 and use this to estimate expected synchrotron emissions from electrons produced via dark matter annihilation. By comparing this to available radio data we review simplifying assumptions used previously for dark matter hunting in these environments and produce novel constraints that are capable of fully ruling out dark matter models proposed to accommodate all the aforementioned excesses barring that of DAMPE. However, we do show that significant constraints can be placed upon the DAMPE parameter space with M31 data. In addition to this we project SKA non-observation constraints for the Reticulum II and Triangulum II dwarf galaxies and find these have potential to rule out cosmic-ray and gamma-ray excess-producing models of dark matter, even when the most conservative assumptions are employed.}

\begin{document}
\maketitle

\section{Introduction}
\label{sec:intro}

Measurements of cosmic-ray fluxes have recently been of some importance in the hunt for the nature of Dark Matter (DM), with many proposed models to account for excess fluxes observed by AMS-02~\cite{amsdocs-ting2013,ams2-antiprotons-aguilar2016}, PAMELA~\cite{pamela-docs,pamela-cr-spectrum2011}, DAMPE~\cite{dampe-ambrosi2017}, HESS\footnote{\url{https://www.mpi-hd.mpg.de/hfm/HESS/}}~\cite{hess-cr-aharonian2008} and Fermi~\cite{fermidocs-atwood2009,fermi-cr-2017}. For an admittedly non-exhaustive list of related works on the topic of DM and cosmic-rays see \cite{dm-cr-bergstrom1999,dm-cr-hooper2003,dm-cr-bringmann2006,dm-cr-hooper2014,dm-cr-cirelli2014,dm-cr-cui2017,dm-cr-cuoco2017a,dm-cr-cuoco2017b} and references therein.  

A recent work \cite{cholis2019} has re-analysed the significance of the observed AMS-02 anti-proton excess, finding much agreement with earlier work on AMS-02 electron-positron fluxes in \cite{dimauro2015}. This accompanied by other recent work in \cite{carena2019} which uses a $60$ GeV neutralino to explain Large Hadron Collider missing energy as well as AMS-02 anti-proton and Galactic Centre GeV gamma-ray excesses. The authors of \cite{cholis2019} demonstrate the robustness of the excess above expected backgrounds to between $5$ and $7$ $\sigma$ confidence intervals, with models that include a DM component in addition to the background being strongly preferred to background-only models. The best-fit region of WIMP parameter space provided by \cite{cholis2019} also evades most existing indirect detection constraints and was shown has some overlap with the favoured region for the GeV galactic-centre gamma-ray excess~\cite{calore2015} and the models used in \cite{dimauro2015,carena2019}. These recent works display a resurgence of models with WIMP masses in the $40$ to $100$ GeV range initially favoured for the Galactic Centre GeV gamma-ray excess that had been somewhat superseded by more mundane astrophysical explanations~\cite{oleary2015,bartels2015,lee2015,brandt2015,gs2016}.
In addition to this, the DArk Matter Particle Explorer (DAMPE) satellite announced an excess in the electron/positron flux around $1.4$ TeV~\cite{dampe-ambrosi2017}. It has been proposed that this could be accounted for by leptophilic DM models via the annihilation of WIMPs and subsequent decay of a leptophilic mediator~\cite{dampedm1,dampedm2,dampeucmh}. A common element of all these models is the required presence of an over-dense clump of DM within a radius $< 1$ kpc in order for the DM models to also satisfy DM cosmological abundance constraints on the annihilation cross-section. For example: \cite{dampedm1} requires, for a WIMP mass range $1.4$ to $1.7$ TeV, an NFW local clump of DM with mass $2 \times 10^5$ M$_{\odot}$ to $10^8$ M$_{\odot}$, positioned at $d \sim 0.1$ kpc from the solar system.

This resurgence of old DM explanations as well as new ones from \cite{dampedm1} can benefit from a multi-frequency and multi-target approach. Where we use the near-universality of DM in cosmic structures to subject DM explanations to scrutiny in situations outside those that suggested the excess in the first place. In particular it has been argued in the literature that radio frequency observations can play a strong role in probing the nature of DM, especially in light of the upcoming Square Kilometre Array (SKA)~\cite{gsp2015,Colafranceso2015,gs2016,Colafrancesco2007,chan2017,chan2019}. This stance has received some experimental backing with recent works such as \cite{regis2014b,regis2017} (for a summary of DM hunting via radio in dwarf galaxies see \cite{gb2019a}). In particular the faintness of diffuse radio emissions around many cosmic structures means they can provide very strong constraints on DM emissions~\cite{regis2014b,gs2016,regis2017}. 

In this work we will focus in particular on nearby galaxies as target environments. This is due to recent work done in \cite{chan2017,chan2019}, which used radio-frequency data on M33 and M31 to produce constraints on viable DM models through synchrotron emissions produced by secondary electrons from DM annihilation. In particular claiming to be able rule out models designed to account for AMS-02 positron~\cite{dimauro2015} and Galactic Centre GeV gamma-ray~\cite{abazajian2016} excesses. We reconsider both the data used in the aforementioned works and additional data from other sources such as \cite{ficarra1985,ned_m31_rice1988,ned_m31_condon2002,ned_m31_jarrett2003,ned_m31_still2009,nvss1998}. Unlike \cite{chan2017,chan2019} our modelling of DM emissions includes all diffusive effects and, by numerical means, is somewhat less approximate in its treatment of DM synchrotron emissions. We demonstrate that the impact of the results in \cite{chan2017,chan2019} is somewhat predicated on their assumption of a constant magnetic field and no diffusion. A radially varying magnetic field produces results that differ noticeably from \cite{chan2019}, while the same occurs for light lepton annihilation channels in M33. We also derive novel constraints under more realistic assumptions in both M31 and M33. These, with the aid of larger data sets, show that radio DM constraints can rule out the entire parameter space for both the GeV gamma-ray~\cite{calore2015}, AMS-02 positron~\cite{dimauro2015}, and anti-proton excesses~\cite{cholis2019} without the dependence on assumptions of substructure boosting, halo density profile, or magnetic field geometry. In addition to this the DM parameter space for the DAMPE excess from \cite{dampedm1} can be partially explored by M31 data. Notably, the constraints found here allow many annihilation channels to be probed to below the thermal relic cross-section~\cite{steigman2012} for WIMP masses up to $100$-$1000$ GeV. In this regard we agree with, and advance upon previous work done on radio-frequency DM hunting in M31 in \cite{egorov2013}.  

This work is structured as follows: sections~\ref{sec:ann} and \ref{sec:emm} detail the formalism for calculating DM radio emissions, section~\ref{sec:halos} displays the properties used for all the DM halos studied, section~\ref{sec:data} lists the radio data used here, sections~\ref{sec:m31}, \ref{sec:m33}, \ref{sec:ex} show our results, and conclusions are drawn in section~\ref{sec:conc}.

\section{Electrons from dark matter annihilation}
\label{sec:ann}

The source function for the production of electrons and positrons (hereafter just referred to as electrons) by WIMP annihilation in a DM halo is taken as
\begin{equation}
Q_e (r,E) = \frac{1}{2}\langle \sigma V\rangle \sum\limits_{f}^{} \td{N^f_e}{E}{} B_f \left(\frac{\rho_{\chi}(r)}{m_{\chi}}\right)^2 \; ,
\end{equation}
where $f$ denotes the annihilation channel, $\langle \sigma V \rangle$ is the velocity-averaged annihilation cross-section, $\td{N^f_i}{E}{}$ is the production spectrum of electrons per annihilation~\cite{ppdmcb1,ppdmcb2}, $B_f$ is the branching fraction of channel $f$, $\rho_{\chi}(r)$ is the DM density at distance $r$ from the halo centre, and $m_{\chi}$ is the WIMP mass.

\section{Synchrotron emission from dark matter annihilation}
\label{sec:emm}

Since electrons can be produced, either as primary or secondary products, in the annihilation of DM particles and magnetic fields are ubiquitous in cosmic structure we can expect synchrotron radiation to result.
The average power of the synchrotron radiation at observed frequency $\nu$ emitted by an electron with energy $E$ in a magnetic field with amplitude $B$ is given by~\cite{longair1994,rybicki1986}
\begin{equation}
P_{synch} (\nu,E,r,z) = \int_0^\pi d\theta \, \frac{\sin{\theta}}{2}2\pi \sqrt{3} r_e m_e c \nu_g F_{synch}\left(\frac{\kappa}{\sin{\theta}}\right) \; ,
\label{eq:power}
\end{equation}
where $m_e$ is the electron mass, $\nu_g = \frac{e B}{2\pi m_e c}$ is the non-relativistic gyro-frequency, $r_e = \frac{e^2}{m_e c^2}$ is the classical electron radius, and the quantities $\kappa$ and $F_{synch}$ are defined as
\begin{equation}
\kappa = \frac{2\nu (1+z)}{3\nu_g \gamma^2}\left[1 +\left(\frac{\gamma \nu_p}{\nu (1+z)}\right)^2\right]^{\frac{3}{2}} \; ,
\end{equation}
with the plasma frequency $\nu_p \propto \sqrt{n_e}$, $\gamma$ as the electron Lorentz factor, and
\begin{equation}
F_{synch}(x) = x \int_x^{\infty} dy \, K_{5/3}(y) \approx 1.25 x^{\frac{1}{3}} \mbox{e}^{-x} \left(648 + x^2\right)^{\frac{1}{12}} \; .
\end{equation}

The emissivity from a population of electrons and positrons with energy spectra $\td{n_{e^-}}{E}{}$ and $\td{n_{e^+}}{E}{}$ is then found via
\begin{equation}
j_{synch} (\nu,r,z) = \int_{m_e}^{m_\chi} dE \, \left(\td{n_{e^-}}{E}{} + \td{n_{e^+}}{E}{}\right) P_{synch} (\nu,E,r,z) \; ,
\label{eq:emm}
\end{equation}
note that in this work $\td{n_{e^-}}{E}{}$ is the equilibrium electron distribution from DM annihilation (see below).
The flux density spectrum within a radius $r$ is then written as
\begin{equation}
S_{synch} (\nu,z) = \int_0^r d^3r^{\prime} \, \frac{j_{synch}(\nu,r^{\prime},z)}{4 \pi (D_L^2+\left(r^{\prime}\right)^2)} \; ,
\label{eq:flux}
\end{equation}
where $D_L$ is the luminosity distance to the target DM halo.

The equilibrium electron distribution is found as a stationary solution to the equation
\begin{equation}
\begin{aligned}
\pd{}{t}{}\td{n_e}{E}{} = & \; \gv{\nabla} \left( D(E,\v{r})\gv{\nabla}\td{n_e}{E}{}\right) + \pd{}{E}{}\left( b(E,\v{r}) \td{n_e}{E}{}\right) + Q_e(E,\v{r}) \; ,
\end{aligned}
\end{equation}
where $D(E,\v{r})$ is the diffusion coefficient, $b(E,\v{r})$ is the energy loss function, and $Q_e(E,\v{r})$ is the electron source function from DM annihilation. In this case, we will work under the simplifying assumption that $D$ and $b$ lack a spatial dependence and thus we will include only average values for magnetic field and gas densities. The solution, when diffusion is negligible, has the form~\cite{Colafrancesco2006}
\begin{equation}
\td{n_e}{E}{} = \frac{1}{b(E)} \int_E^{m_\chi} \, dE^{\prime} \, Q_e (r, E^{\prime}) \; .
\end{equation}
When diffusion is not negligible, as in dwarf galaxies, a spherically symmetric solution can be found~\cite{baltz1999,baltz2004,Colafrancesco2006,Colafrancesco2007}
\begin{equation}
\td{n_e}{E}{} (r,E) = \frac{1}{b(E)}  \int_E^{M_\chi} d E^{\prime} \, G(r,E,E^{\prime}) Q (r,E^{\prime}) \; ,
\end{equation}
by means of a Green's function $G(r,E,E^{\prime})$. This function is expressed as
\begin{equation}
\begin{aligned}
G(r,E,E^{\prime}) = & \frac{1}{\sqrt{4\pi\Delta v}} \sum_{n=-\infty}^{\infty} (-1)^n \int_0^{r_h} d r^{\prime} \; \frac{r^{\prime}}{r_n} \\ & \times \left( \exp\left(-\frac{\left(r^{\prime} - r_n\right)^2}{4\Delta v}\right) - \exp\left(-\frac{\left(r^{\prime} + r_n\right)^2}{4\Delta v}\right) \right)\frac{Q(r^{\prime})}{Q(r)} \; ,
\end{aligned}
\end{equation}
where the sum runs over the indices of a set of image charges at positions given by $r_n = (-1)^n r + 2 n r_h$. The radius $r_h$ is the maximum radius we consider diffusion of particles out to. In this work we will use $r_h = 2 R_{vir}$ with $R_{vir}$ being the virial radius of the halo in question. Then $\Delta v$ is defined
\begin{equation}
\Delta v =  v(u(E)) - v(u(E^{\prime})) \; ,
\end{equation}
with
\begin{equation}
\begin{aligned}
v(u(E)) = & \int_{u_{min}}^{u(E)} dx \; D(x) \; , \\
u (E) = & \int_E^{E_{max}} \frac{dx}{b(x)} \; . \\ 
\end{aligned}
\end{equation}

In order to complete this solution we then need to define the diffusion and energy-loss functions. For $D(E)$ we follow \cite{Colafrancesco1998} so that
\begin{equation}
D(E) = \frac{1}{3}c r_L (E) \frac{\overline{B}^2}{\int^{\infty}_{k_L} dk P(k)} \; ,
\end{equation}
where $\overline{B}$ is the average magnetic field, $r_L$ is the Larmour radius of a relativistic particle with energy $E$ and charge $e$ and $k_L = \frac{1}{r_L}$. This combined with the requirement that
\begin{equation}
\int^{\infty}_{k_0} dk P(k) = \overline{B}^2 \; ,
\end{equation}
where $k_0 = \frac{1}{d_0}$, with $d_0$ being the smallest scale on which the magnetic field is homogeneous, yields the final form
\begin{equation}
D(E) = D_0 \left(\frac{d_0}{1 \; \mbox{kpc}}\right)^{\frac{2}{3}} \left(\frac{\overline{B}}{1 \; \mu\mbox{G}}\right)^{-\frac{1}{3}} \left(\frac{E}{1 \; \mbox{GeV}}\right)^{\frac{1}{3}}  \; , \label{eq:diff}
\end{equation}
where $D_0 = 3.1\times 10^{28}$ cm$^2$ s$^{-1}$.

The energy loss function is defined by~\cite{Colafrancesco2006,Colafrancesco2007}
\begin{equation}
\begin{aligned}
b(E) = b_{IC} E^2 (1+z)^4 + b_{sync} E^2 \overline{B}^2 \; + b_{Coul} \overline{n} \left(1 + \frac{1}{75}\log\left(\frac{\gamma}{\overline{n}}\right)\right) + b_{brem} \overline{n} \left( \log\left(\frac{\gamma}{\overline{n}}\right) + 0.36 \right) \;,
\end{aligned}
\label{eq:loss}
\end{equation}
where $\overline{n}$ is the average gas density in the halo and is given in cm$^{-3}$, $\overline{B}$ is the average magnetic field in $\mu$G, $E$ is the electron energy in GeV, while $b_{IC}$, $b_{synch}$, $b_{col}$, and $b_{brem}$ are the inverse Compton, synchrotron, Coulomb and bremsstrahlung energy loss factors. These are given, in units of $10^{-16}$ GeV s$^{-1}$, by $6.08$/$0.25$, $0.0254$, $6.13$, and $1.51$ respectively. The two values of $b_{IC}$ represent the cases of a galactic inter-stellar radiation field, and that of the CMB only respectively.

\section{Dark Matter Halos}
\label{sec:halos}

\subsection{M31}
For the M31 DM halo we will follow the studies made in \cite{tamm2012} and make use of three different halo profiles: Navarro-Frenk-White (NFW)~\cite{nfw1996}, Burkert~\cite{burkert1995}, and Einasto~\cite{einasto1968} summarised in Eq.~(\ref{eq:density}).
\begin{equation}
\begin{aligned}
\rho_{nfw}(r)=\frac{\rho_{s}}{\frac{r}{r_s}\left(1+\frac{r}{r_s}\right)^{2}} \; , \\
\rho_{burk}(r)=\frac{\rho_s}{\left(1+ \frac{r}{r_s}\right)\left(1+\left[\frac{r}{r_s}\right]^2\right)} \; , \\
\rho_{ein}(r)=\rho_{s} \exp\left[-\frac{2}{\alpha} \left(\left[\frac{r}{r_s}\right]^{\alpha} - 1\right)\right] \; .
\end{aligned}
\label{eq:density}
\end{equation}
Note that $r_s$ and $\rho_s$ have different physical roles in each of the profiles, however, we label them in this manner for convenience of presentation in Table~\ref{tab:m31} which is sourced from the modelling of \cite{tamm2012}. The distance to the halo centre is taken to be 770 kpc.
\begin{table}[htbp]
	\centering
	\begin{tabular}{|c|c|c|c|c|}
		\hline
		Profile & $\rho_s$ (M$_{\odot}$ pc$^{-3}$) & $r_s$ (kpc) & $M_{vir}$ ($10^{10}$ M$_{\odot}$) & $R_{vir}$ (kpc) \\
		\hline 
		NFW & $1.10 \pm 0.18 \times 10^{-2}$ & $16.5 \pm 1.5$ & $104$ & $207$ \\
		Burkert & $3.68 \pm 0.40 \times 10^{-2}$ & $9.06 \pm 0.53$ & $79$ & $189$ \\
		Einasto ($\alpha = 0.17$) & $8.12 \pm 0.16 \times 10^{-6}$ & $17.44$ & $113$ & $213$ \\
		\hline
	\end{tabular}
\caption{M31 Density Profile Properties from \cite{tamm2012}. $M_{vir}$ and $R_{vir}$ are the virial mass and radius.}
\label{tab:m31}
\end{table}
For the magnetic field and the gas distributions we will follow the model of \cite{ruiz-granados2010} and take it so that our magnetic field strength is $B = 4.6 \pm 1.2$ $\mu$G at $r = 14$ kpc, with a more general profile following
\begin{equation}
B(r) = \frac{4.6 r_1 + 64}{r_1 + r} \, \mu\mathrm{G} \; ,
\end{equation}
where $r_1 = 200$ kpc is taken to follow the more conservative value from fitting in \cite{ruiz-granados2010}. 
We take the gas density to be given by an exponential profile
\begin{equation}
n_e (r) = n_0 \exp\left(-\frac{r}{r_d}\right) \; ,
\end{equation}
where $n_0 = 0.06$ cm$^{-3}$ is the central density~\cite{beckm31}, and $r_d \approx 5$ kpc is the disk scale radius fitted by \cite{ruiz-granados2010}. As argued in \cite{chan2019} spatial diffusion of secondary electrons can be neglected in this magnetic field environment, as its time-scale is far longer than that for energy losses (this was verified by calculation for all presented cases).

\subsection{M33}
For M33 we will implement both Burkert and NFW density profiles. In the case of NFW we will use $M_{vir} = 5.4 \pm 0.6\times 10^{11}$ M$_{\odot}$, and $r_{s} = 15.3 \pm 0.56$ kpc to fully specify the density profile~\cite{fune2016}. For a Burkert form we will use $M_{vir} = 3 \pm 0.8 \times 10^{11}$ M$_{\odot}$ and $r_s = 9.6 \pm 0.5$ kpc~\cite{fune2016}. The distance to the halo centre is taken to be 840 kpc.
The magnetic field radial profile will be assumed to follow an exponential form (as this is common for spiral galaxies~\cite{beck2015}) and we will take $B_0$ such that the average field within $7.5$ kpc is $8.1 \pm 0.5$ $\mu$G following~\cite{chan2017} based on the results from \cite{berkhuijsen2013}. The scale-length of the magnetic field will be taken to be $\approx 5$ kpc following arguments in \cite{beck2015} that is $\approx 3.8 r_d$ where $r_d$ is the scale-length of the baryonic matter distribution. We will consider diffusion with a minimal coherence length of $50$ pc for the turbulent magnetic field to match spiral galaxies of this size~\cite{beck2015}. For the purposes of energy-loss and diffusion we will consider the average magnetic field to be given by $8.1$ $\mu$G to produce more conservative results. The gas central density is taken as $0.03$ cm$^{-3}$~\cite{beckm31} and is taken to follow an exponential radial profile, with scale-radius $r_d = 1.2 \pm 0.2$ kpc following \cite{regan1994}.   

\subsection{Dwarf spheroidals}
For Reticulum II and Triangulum II dwarf galaxies we will assume a Burkert density profile (as this is favoured for dwarf galaxies~\cite{walker2009,adams2014}) normalised to a kinematically determined J-factor $J = 10^{19.6}$ GeV$^2$ cm$^{-5}$ for 0.5 degrees~\cite{bonnivard2015} for Reticulum II and for Triangulum II we take $J = 10^{21.03}$ GeV$^2$ cm$^{-5}$ \cite{genina2016}. $J$ is given by
\begin{equation}
J (\Delta \Omega, l) = \int_{\Delta \Omega}\int_{l} \rho^2 (\v{r}) dl^{\prime}d\Omega^{\prime} \; , \label{eq:jfactor}
\end{equation}
with the integral being extended over the line of sight $l$, and $\Delta \Omega$ is the observed solid angle. 
The magnetic field and electron densities will both be assumed to follow exponential distributions with a scale radius $r_d = 15$ pc in Reticulum II and $r_d = 35.68$ pc in Triangulum II. These scales being chosen as they are the stellar half-light radii of these targets~\cite{bechtol2015,koposov2015,laevens2015}. We assume the magnetic fields in both dwarf galaxies have Kolmogorov turbulence spectra with a minimum coherence length of $1$ pc. The central values for magnetic field and electron density will be assumed, in both cases, to be $B_0 \approx 1$ $\mu$G and $n_0 \approx 10^{-6}$ cm$^{-3}$. Note that these are chosen to be slightly conservative following arguments made in \cite{regis2014b,regis2017}.

\subsection{The effect of substructure}
The expected presence of sub-halos within a DM halo can produce a substantial increase in fluxes resulting from DM annihilation~\cite{strigari2007,gs2016}. A common approach in the literature is to calculate a ``boosting factor" by which the presence of sub-halos will increase the DM annihilation flux. In order to compare our results directly to \cite{chan2017,chan2019} we will employ the same boosting factors following the calculation method in \cite{moline2017}. These are, for M31 and M33 respectively, $5.28$ and $4.86$. We will employ no boosting from substructure in the case of dwarf galaxies as this is not expected to be significant due to their low mass.   

\section{Flux data sets}
\label{sec:data}
This work makes uses of several sets of flux data. In the case of M31 we use the integrated flux found by \cite{chan2019}, between $4.6$ and $5$ GHz with Region of Interest (ROI) covering a 40 kpc radius, as well as lower frequency fluxes integrated over a map of the target taken from the Nasa Extra-galactic Database (NED)\footnote{\url{http://ned.ipac.caltech.edu/}} and sourced from \cite{ficarra1985,ned_m31_rice1988,ned_m31_condon2002,ned_m31_jarrett2003,ned_m31_still2009} which will be compared to DM radio spectra integrated out to an ROI radius chosen to meet each case. We note that the low frequency 408 MHz point \cite{ficarra1985} has a significant impact on the results. However, it does not have a quoted ROI but it does have a maximum observable angular scale of at least $70$ arcminutes according to the configuration quoted in \cite{ficarra1985}. However, due to the uncertainty we will use three cases for this data point: a pessimistic one with a 5 arcminute ROI, a median case of 20 arcminutes (4.5 kpc), and an optimistic one with a 50 arcminute (11 kpc) ROI. In the case of \cite{ned_m31_rice1988,ned_m31_jarrett2003} the observations were done with the MIPS/IRAC instruments on Spitzer with $\gtrsim 50$~\cite{ned_m31_rice1988}/51~\cite{ned_m31_jarrett2003} arcminute angular radii, so we use a 50 arcminute ROI in these cases. The observations for \cite{ned_m31_condon2002} come from the VLA in mode D, with a largest recoverable scale of 16 arcminutes, therefore we use 15 arcminutes as the ROI in this case. Additionally, \cite{ned_m31_still2009} was conducted with the Effelsberg telescope so we need not consider the largest angular scale, so we use the 40 kpc ROI from \cite{chan2019} due to the similar frequency range and lack of additional sources at this frequency within the ROI~\cite{chan2019}. We summarise the M31 data in Table~\ref{tab:m31-flux}.

\begin{table}[htb]
	\centering
	\begin{tabular}{|c|c|c|c|c|}
		\hline
		$\nu$ (MHz) & $S(\nu)$ (Jy) & $\Delta S(\nu)$ (Jy) & ROI & Ref\\
		\hline
		408.0 &  0.22  & 0.02 & 5 - 50 arcmin & \cite{ficarra1985} \\
		1.4e3 & 8.6 & Not given & $\approx 15$ arcmin & \cite{ned_m31_condon2002} \\
		4.8e3  & 1.863 &  Not given & 40 kpc & \cite{ned_m31_still2009} \\
		3.0e6 &  2928.40 &  439.26 & $\approx 50$ arcmin & \cite{ned_m31_rice1988} \\
		5e6 & 536.18 & 80.427 & $\approx 50$ arcmin & \cite{ned_m31_rice1988} \\ 
		1.2e7 & 107.71 &  16.1565 & $\approx 50$ arcmin & \cite{ned_m31_rice1988} \\
		1.38e8 & 268.0 & 4.25  & $\approx 50$ arcmin & \cite{ned_m31_jarrett2003} \\ 
		1.82e8 &  3.14e2 & 4.96   & $\approx 50$ arcmin & \cite{ned_m31_jarrett2003} \\ 
		2.4e8 & 231.0 & 3.44  & $\approx 50$ arcmin & \cite{ned_m31_jarrett2003} \\ 
		\hline
	\end{tabular}
	\caption{M31 flux data points used. See text for details.}
	\label{tab:m31-flux}
\end{table}

For M33 we employ two data sets. One using just upper limits from \cite{chan2017} integrated 7.5 kpc from the centre M33. The other uses the data from \cite{chan2017} as well as results from \cite{nvss1998}. Note that, when comparing M33 DM spectra to \cite{nvss1998} we only integrate over a 1.5 arcminute area around the halo centre (this small area will amplify the need to consider diffusive effects for this data in M33).

In the case of dwarf galaxies we will find non-observation constraints by using the sensitivity profile of SKA-1~\cite{ska2012} at the $2\sigma$ confidence interval with $50$ hours of observation time. 

\section{Constraints from M31}
\label{sec:m31}
In each case we will find the smallest value of $\langle \sigma V \rangle$ that is excluded by the data at a confidence level of $2\sigma$. As a reference case we will also plot the constraints on the $b\bar{b}$ and $\tau^+\tau^-$ channels from Fermi-LAT studies of dwarf galaxies~\cite{Fermidwarves2015,Fermidwarves2016}.

Note that we use the data from \cite{chan2019} with an ROI of 40 kpc. To compare to this we consider two cases, one where all of the data \cite{ficarra1985,ned_m31_rice1988,ned_m31_condon2002,ned_m31_jarrett2003,ned_m31_still2009} is considered to have an ROI of 40 kpc, and the other where \cite{ficarra1985} has an ROI of 5 arcminutes while the others are unchanged. 

\begin{figure}[htbp]
	\centering
	\resizebox{0.49\hsize}{!}{\includegraphics{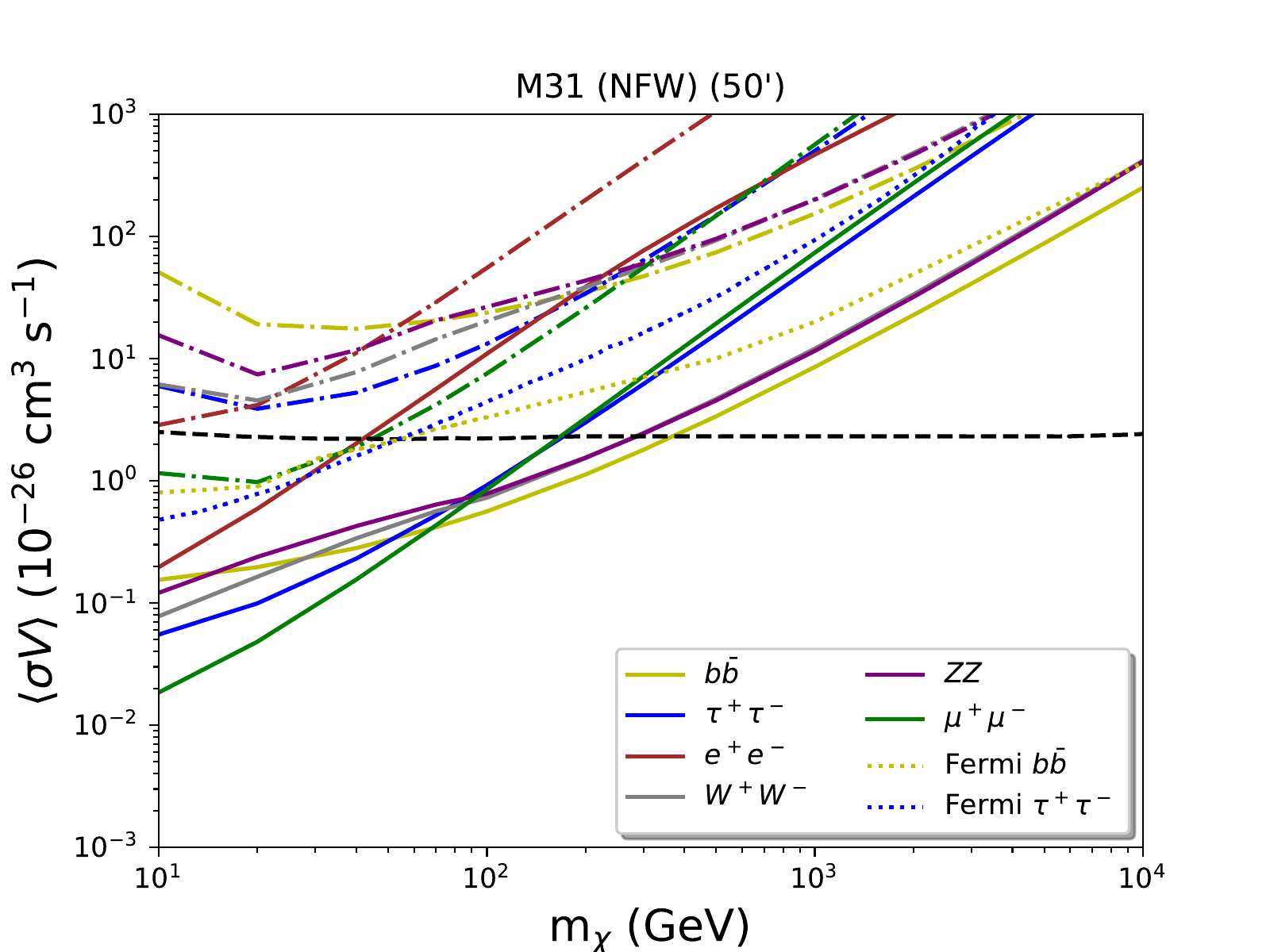}}
	\resizebox{0.49\hsize}{!}{\includegraphics{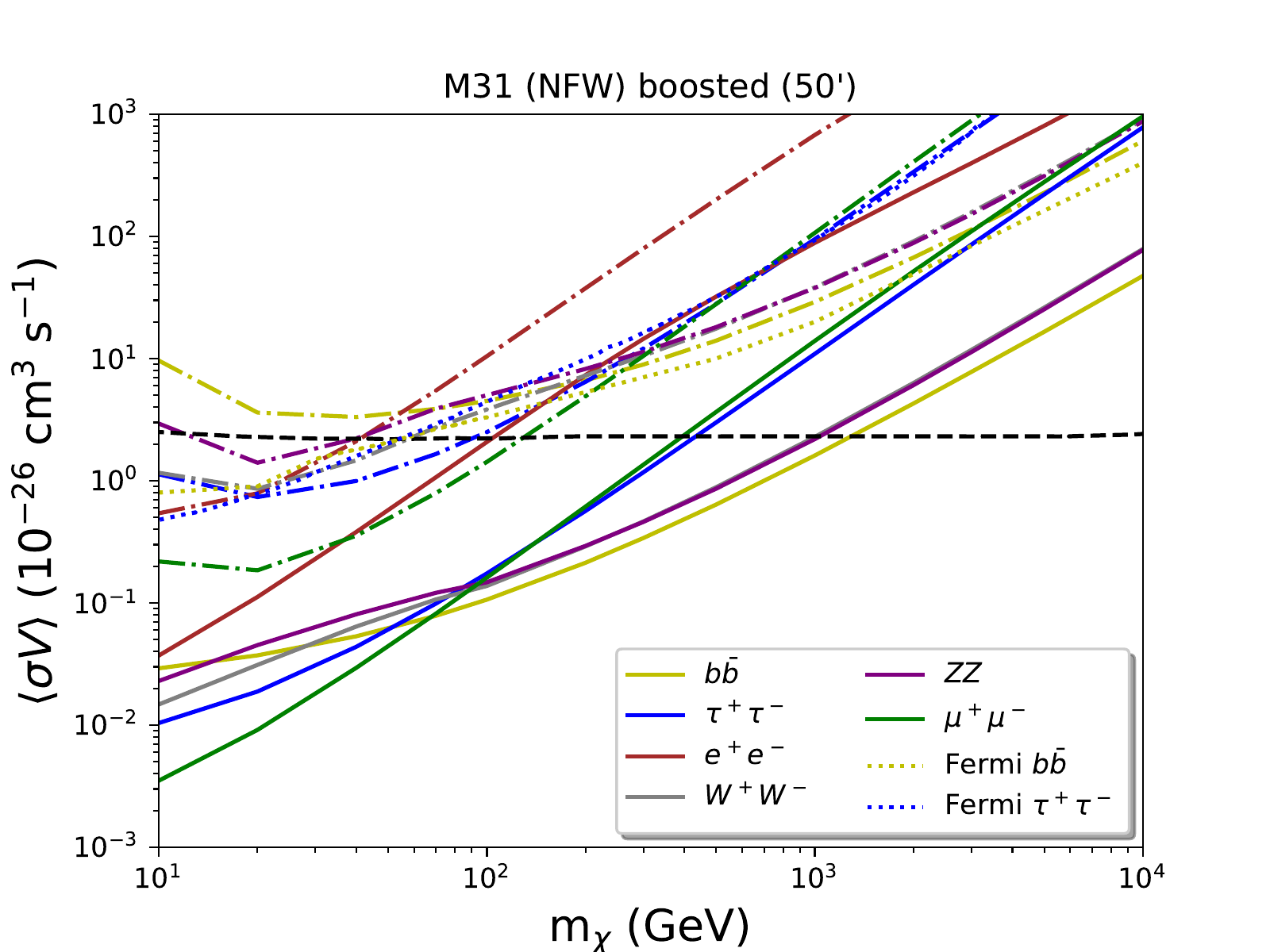}}
	\caption{M31 cross-section upper limits at $2\sigma$ confidence level for NFW halos with a 50 arcminute ROI at 408 MHz. Left: unboosted. Right: boost factor $5.28$. The black dashed line shows the thermal relic cross-section~\cite{steigman2012}. The dotted lines show Fermi-LAT dwarf galaxy limits from \cite{Fermidwarves2016}. The solid lines show M31 results from this work for various annihilation channels using data points listed in Section~\ref{sec:data}, while dash-dotted lines do the same for the integrated flux from \cite{chan2019}.}
	\label{fig:m31-nfw}
\end{figure}

\begin{figure}[htbp]
	\centering
	\resizebox{0.49\hsize}{!}{\includegraphics{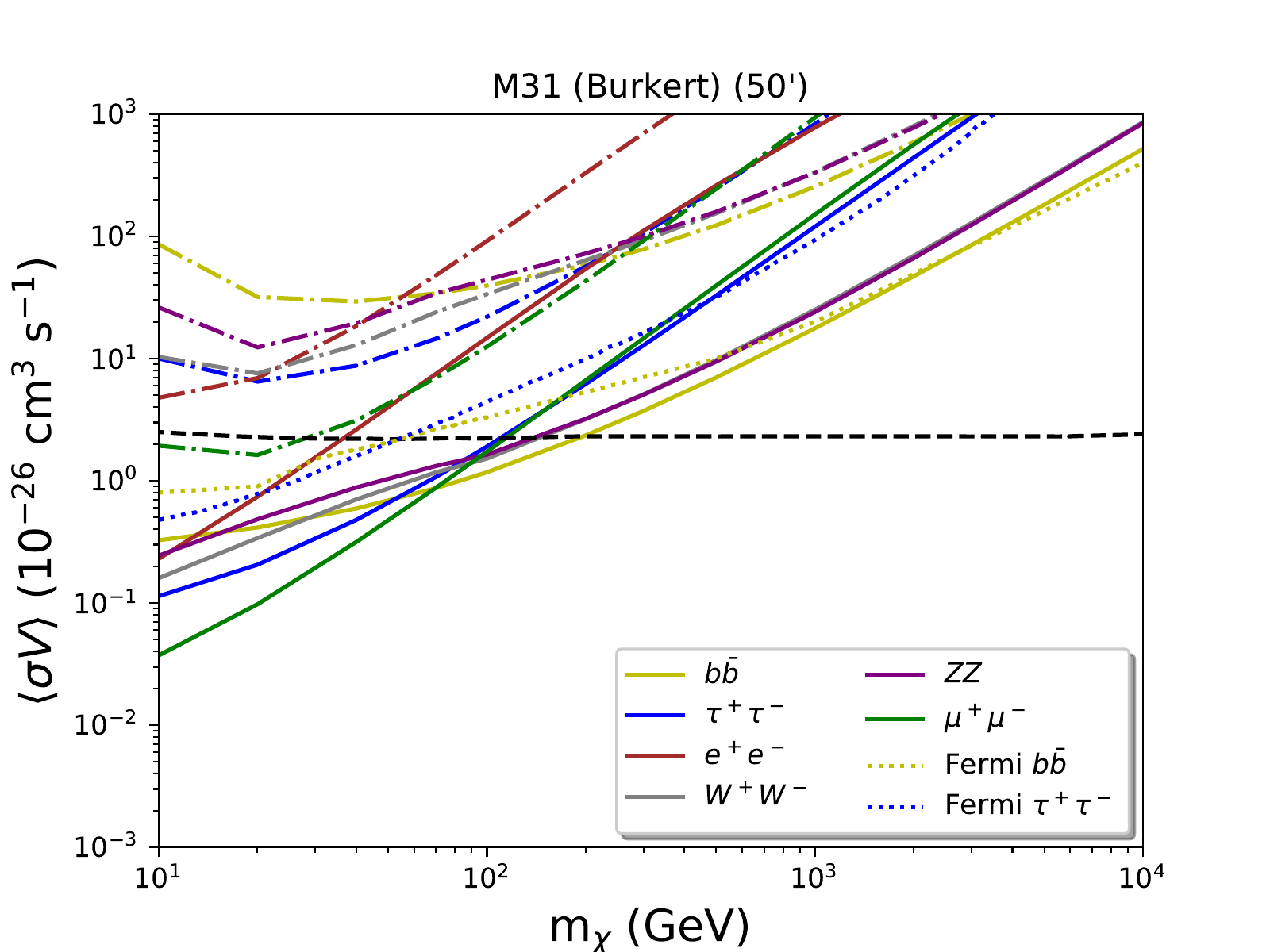}}
	\resizebox{0.49\hsize}{!}{\includegraphics{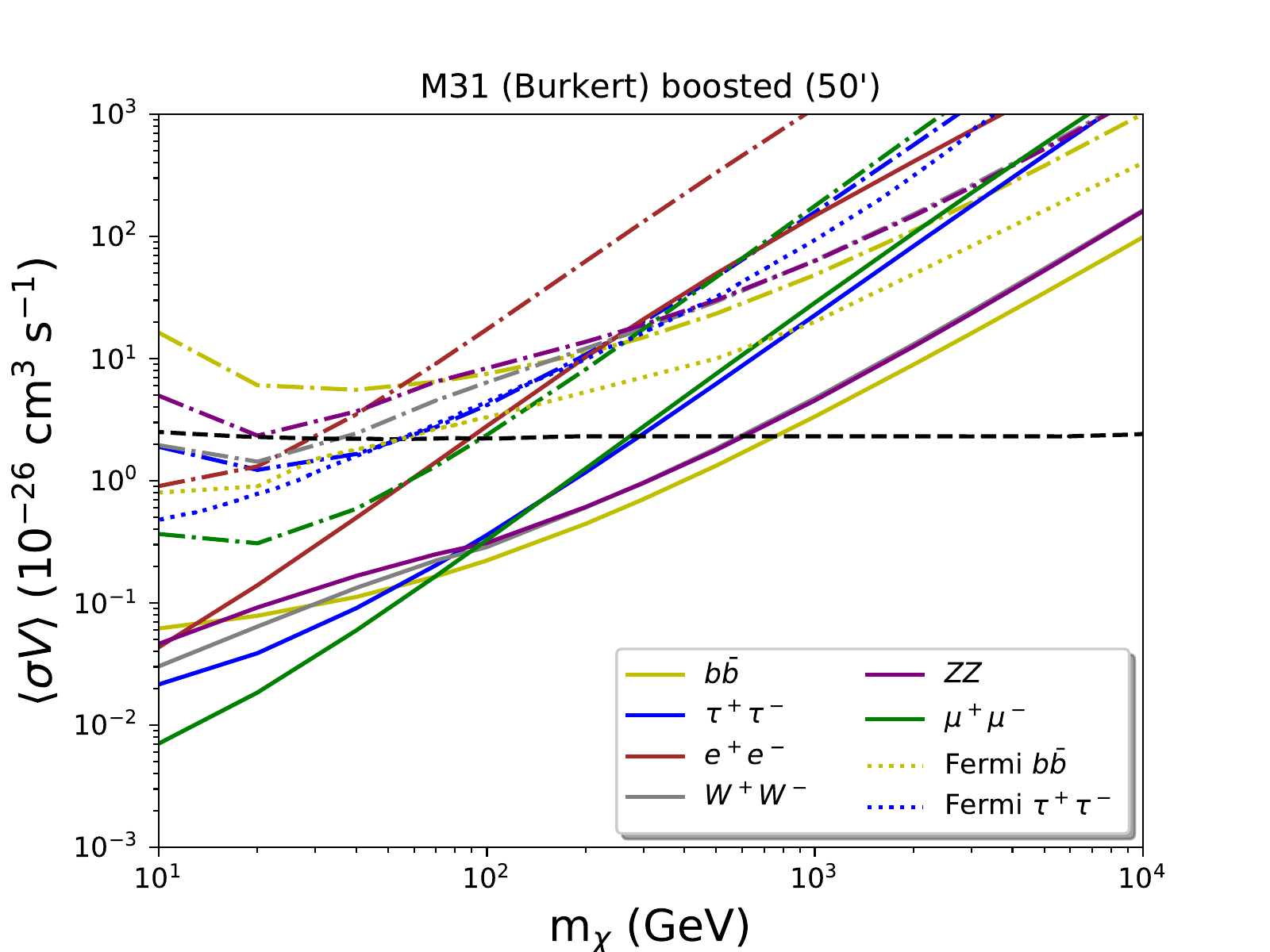}}
	\caption{M31 cross-section upper limits at $2\sigma$ confidence level for Burkert halos with a 50 arcminute ROI at 408 MHz. Left: unboosted. Right: boost factor $5.28$. The black dashed line shows the thermal relic cross-section~\cite{steigman2012}. The dotted lines show Fermi-LAT dwarf galaxy limits from \cite{Fermidwarves2016}. The solid lines show M31 results from this work for various annihilation channels using data points listed in Section~\ref{sec:data}, while dash-dotted lines do the same for the integrated flux from \cite{chan2019}.}
	\label{fig:m31-bur}
\end{figure}

\begin{figure}[htbp]
	\centering
	\resizebox{0.49\hsize}{!}{\includegraphics{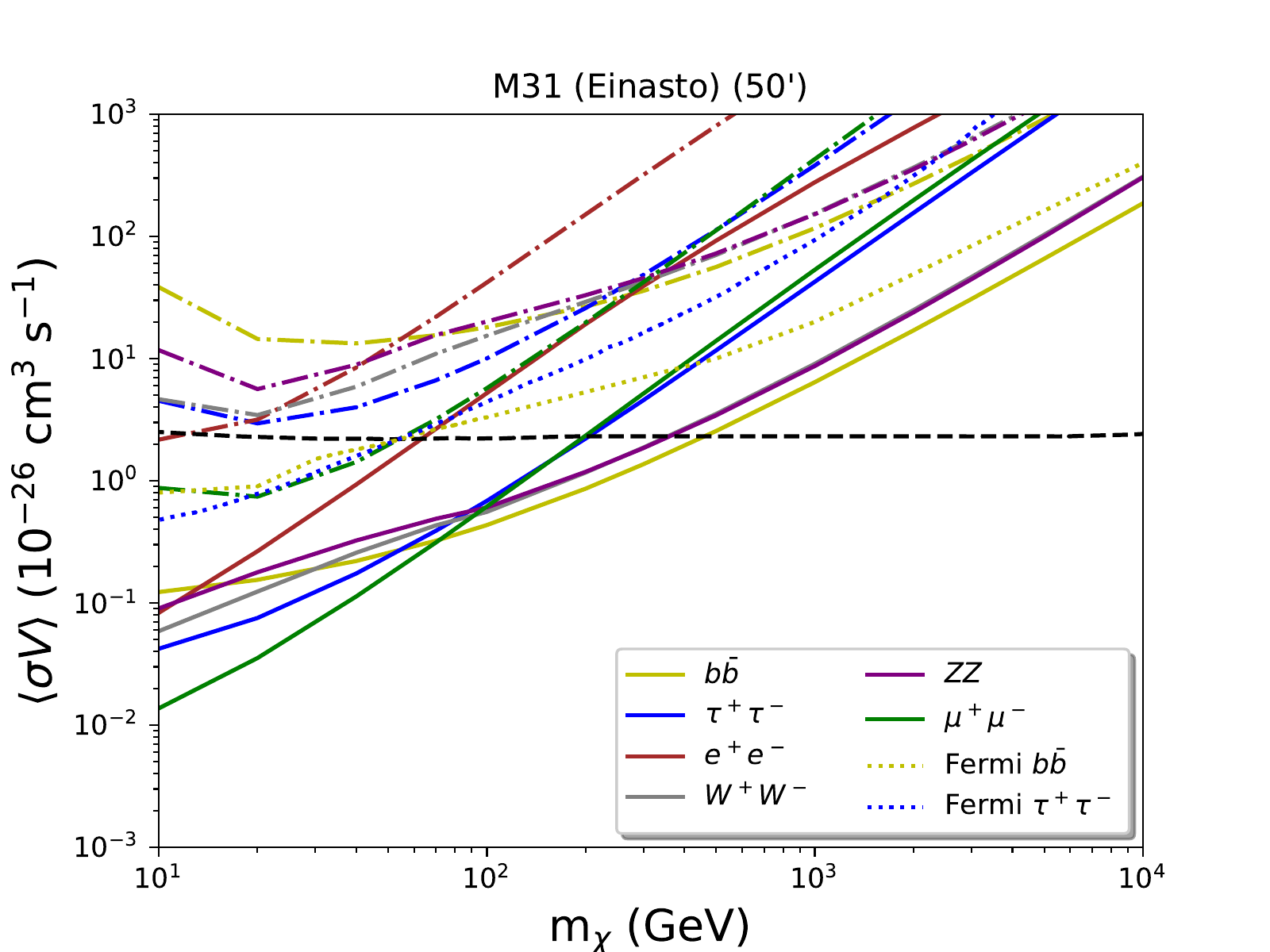}}
	\resizebox{0.49\hsize}{!}{\includegraphics{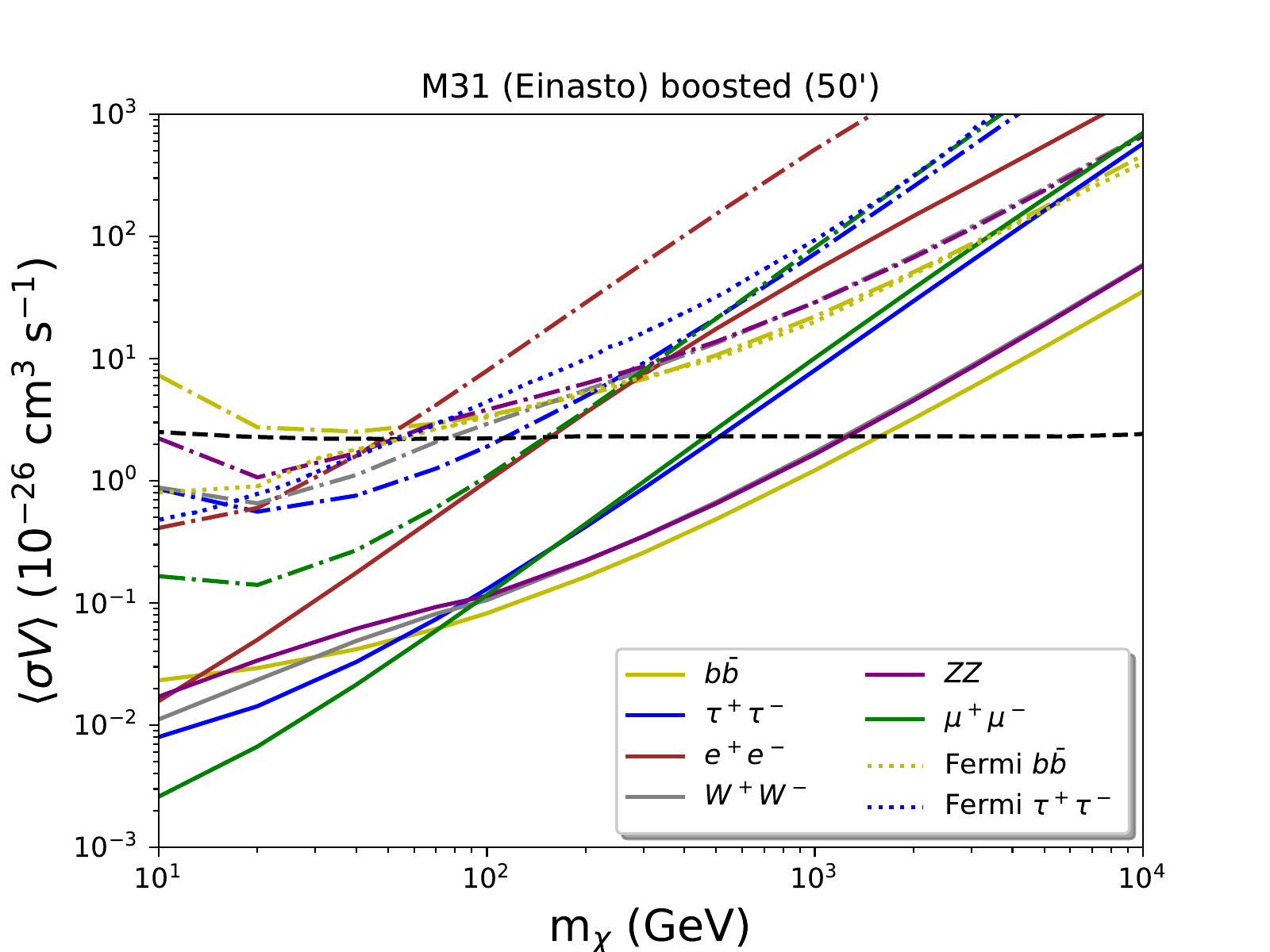}}
	\caption{M31 cross-section upper limits at $2\sigma$ confidence level for Einasto halos with a 50 arcminute ROI at 408 MHz. Left: unboosted. Right: boost factor $5.28$. The black dashed line shows the thermal relic cross-section~\cite{steigman2012}. The dotted lines show Fermi-LAT dwarf galaxy limits from \cite{Fermidwarves2016}. The solid lines show M31 results from this work for various annihilation channels using data points listed in Section~\ref{sec:data}, while dash-dotted lines do the same for the integrated flux from \cite{chan2019}.}
	\label{fig:m31-ein}
\end{figure}

In Figs.~\ref{fig:m31-nfw}, \ref{fig:m31-bur}, and \ref{fig:m31-ein} we display the constraints produced by comparing predicted fluxes to aforementioned radio data for NFW, Burkert, and Einasto halos respectively. These figures use a 50 arcminute (or $\approx 11$ kpc) ROI at 408 MHz. The left panel of each figure displays a case with no halo boosting, comparing the quark channel in this case with the Fermi-LAT dwarf spheroidal limits we can see that there is almost an order of magnitude advantage over Fermi-LAT in the NFW case (solid lines showing the limits from Section~\ref{sec:data} including more data than just \cite{chan2019}) with Burkert and Einasto profiles respectively reducing and increasing the advantage over Fermi-LAT by a factor of 2 compared to NFW. This is significant as limits derived here for the differential case are similar to those reported for the integrated flux between 4.6 and 5 GHz in \cite{chan2019}. However, the integrated flux results found in this work (dot-dashed lines) are significantly weaker than those reported by \cite{chan2019} making use of the same data, being around an order of magnitude weaker than those shown for Fermi-LAT (comparing b-quark channels again). Significantly, the extension of our data set down to 408 MHz makes limits on low-mass WIMPs much stronger than when using the $4.6$ to $5$ GHz data only. In the right-hand column of these figures we use the same boosting factor as reported for M31 in \cite{chan2019} of $5.28$, in this case the b-quark channel results for the radio data from \cite{chan2019} are only reaching parity with Fermi-LAT for WIMP masses $> 100$ GeV for an NFW density profile (contrary to the results in \cite{chan2019} where the exceed Fermi-LAT at all masses), the Burkert case remains roughly a factor of 2 weaker than Fermi-LAT for the same mass range. However, the Einasto case with boosting and \cite{chan2019} radio data surpasses Fermi-LAT for masses $> 30$ GeV. 

These results suggest that the approximations used in \cite{chan2019} do not in fact provide an accurate conservative estimate for DM constraints, these being around an order of magnitude stronger than those found here for the same integrated flux data. The only major points of difference between this work and \cite{chan2019} is the use of differing magnetic field profiles and in our consideration of more energy-loss processes than just synchrotron emission (the latter likely causes around a factor of 3 reduction in DM radio flux~\cite{siffert2011}). In this work we allow the field to vary radially following \cite{ruiz-granados2010} while \cite{chan2019} employs a constant magnetic field with $5$ $\mu$G magnitude. This implies that the argument made in \cite{chan2017,chan2019}, being that the likely higher central value of the magnetic field strength justifies the use of a flat profile, does not hold up in practice. As demonstrated by the fact that the central value for the magnetic field strength used here for M31 is $\gtrsim 10$ $\mu$G, fully compatible with the values argued for in \cite{chan2019}. Although \cite{chan2019} uses the halo profile from \cite{sofue2015} we find that this produces a synchrotron flux smaller by a factor of $\approx 1.2$ (with our magnetic field model), so it will not account for the differences found here. 

While our above analysis has focussed on b-quark channels for the purpose of simple benchmarking against Fermi-LAT, we note that the other annihilation channels display very strong results as well.
We see in figs.~\ref{fig:m31-nfw}, \ref{fig:m31-bur}, and \ref{fig:m31-ein} that the differential constraints (solid lines) are extremely strong, allowing the leptonic annihilation channels to be constrained below the expected thermal relic cross-section all the way up to $\sim 100$ GeV WIMP mass. However, other channels produce more striking results, with constraints below the thermal relic level out to $\sim 1000$ GeV WIMP masses regardless of choice of halo density profile or the use of substructure boosting factors.


\begin{figure}[htbp]
	\centering
	\resizebox{0.49\hsize}{!}{\includegraphics{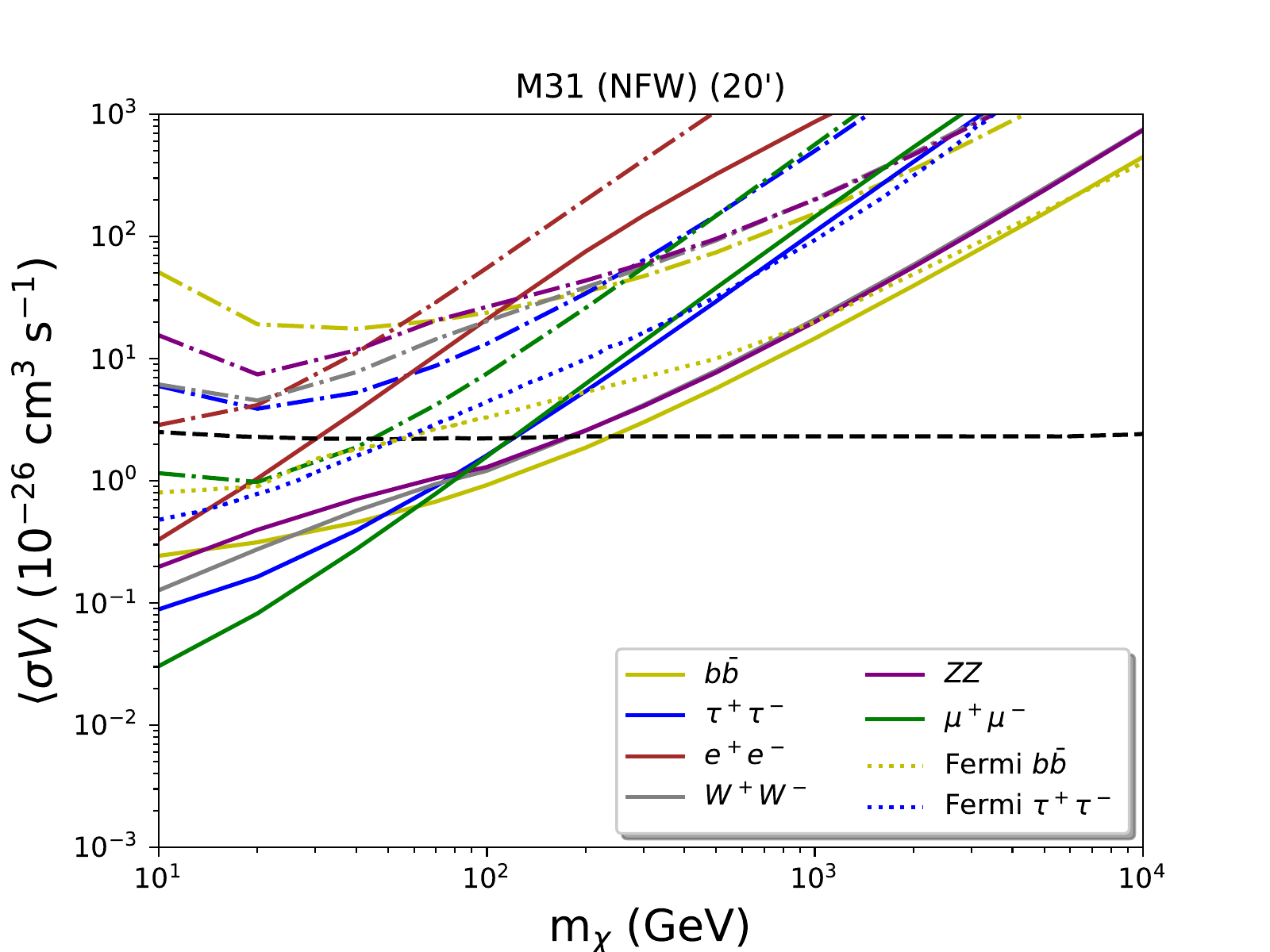}}
	\resizebox{0.49\hsize}{!}{\includegraphics{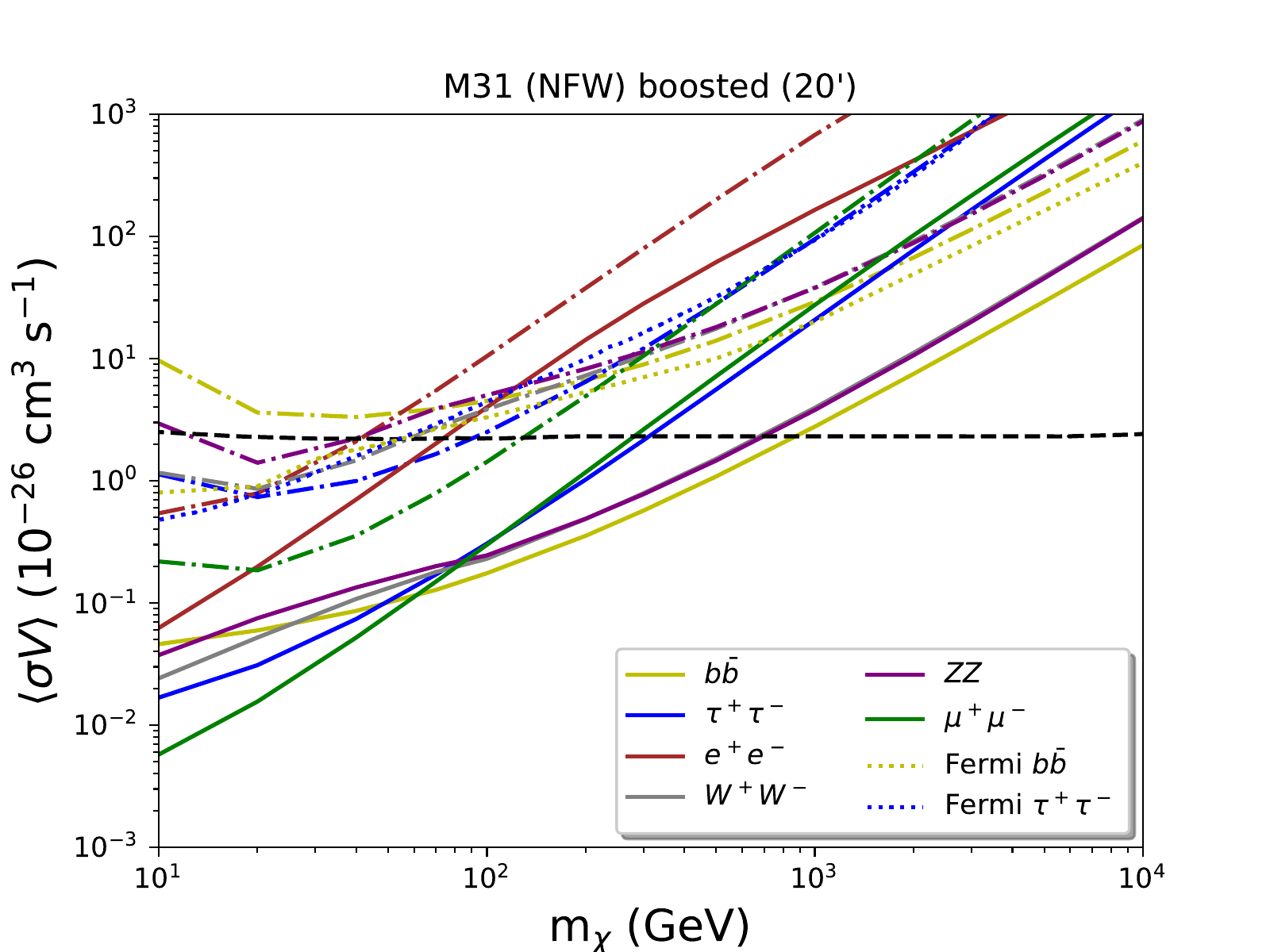}}
	\caption{M31 cross-section upper limits at $2\sigma$ confidence level for NFW halos with a 20 arcminute ROI at 408 MHz. Left: unboosted. Right: boost factor $5.28$. The black dashed line shows the thermal relic cross-section~\cite{steigman2012}. The dotted lines show Fermi-LAT dwarf galaxy limits from \cite{Fermidwarves2016}. The solid lines show M31 results from this work for various annihilation channels using data points listed in Section~\ref{sec:data}, while dash-dotted lines do the same for the integrated flux from \cite{chan2019}.}
	\label{fig:m31-nfw-med}
\end{figure}

\begin{figure}[htbp]
	\centering
	\resizebox{0.49\hsize}{!}{\includegraphics{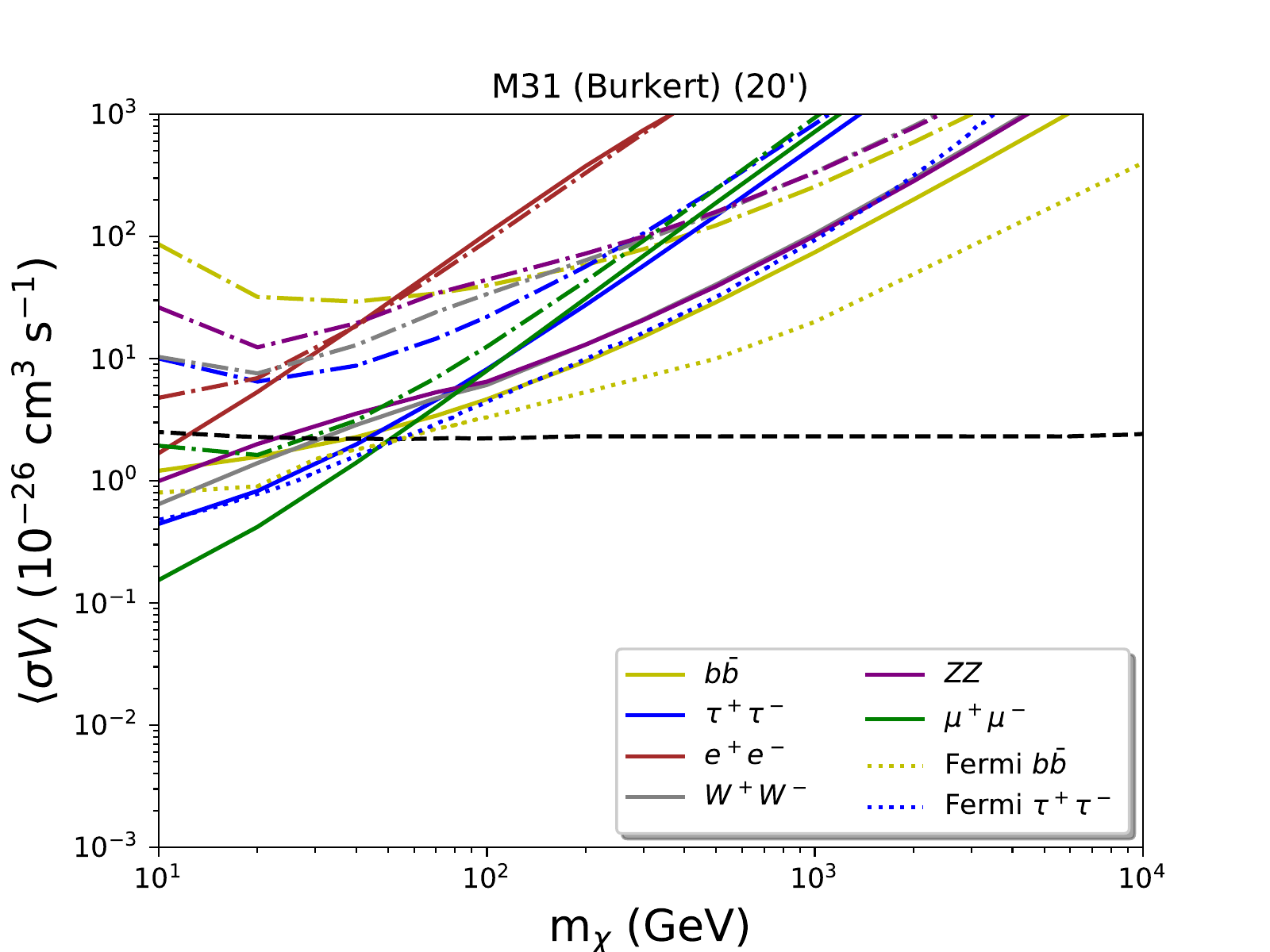}}
	\resizebox{0.49\hsize}{!}{\includegraphics{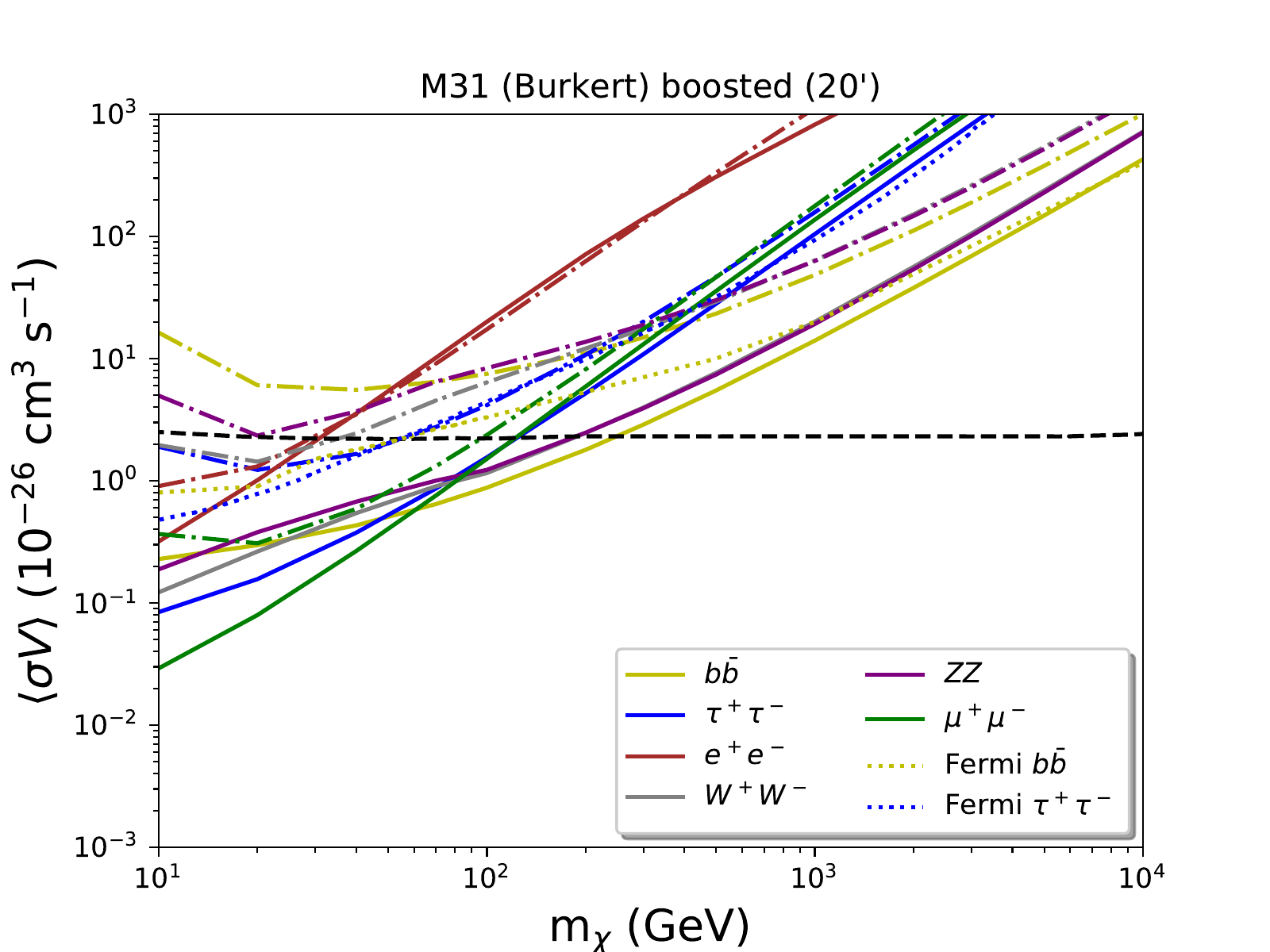}}
	\caption{M31 cross-section upper limits at $2\sigma$ confidence level for Burkert halos with a 20 arcminute ROI at 408 MHz. Left: unboosted. Right: boost factor $5.28$. The black dashed line shows the thermal relic cross-section~\cite{steigman2012}. The dotted lines show Fermi-LAT dwarf galaxy limits from \cite{Fermidwarves2016}. The solid lines show M31 results from this work for various annihilation channels using data points listed in Section~\ref{sec:data}, while dash-dotted lines do the same for the integrated flux from \cite{chan2019}.}
	\label{fig:m31-bur-med}
\end{figure}

\begin{figure}[htbp]
	\centering
	\resizebox{0.49\hsize}{!}{\includegraphics{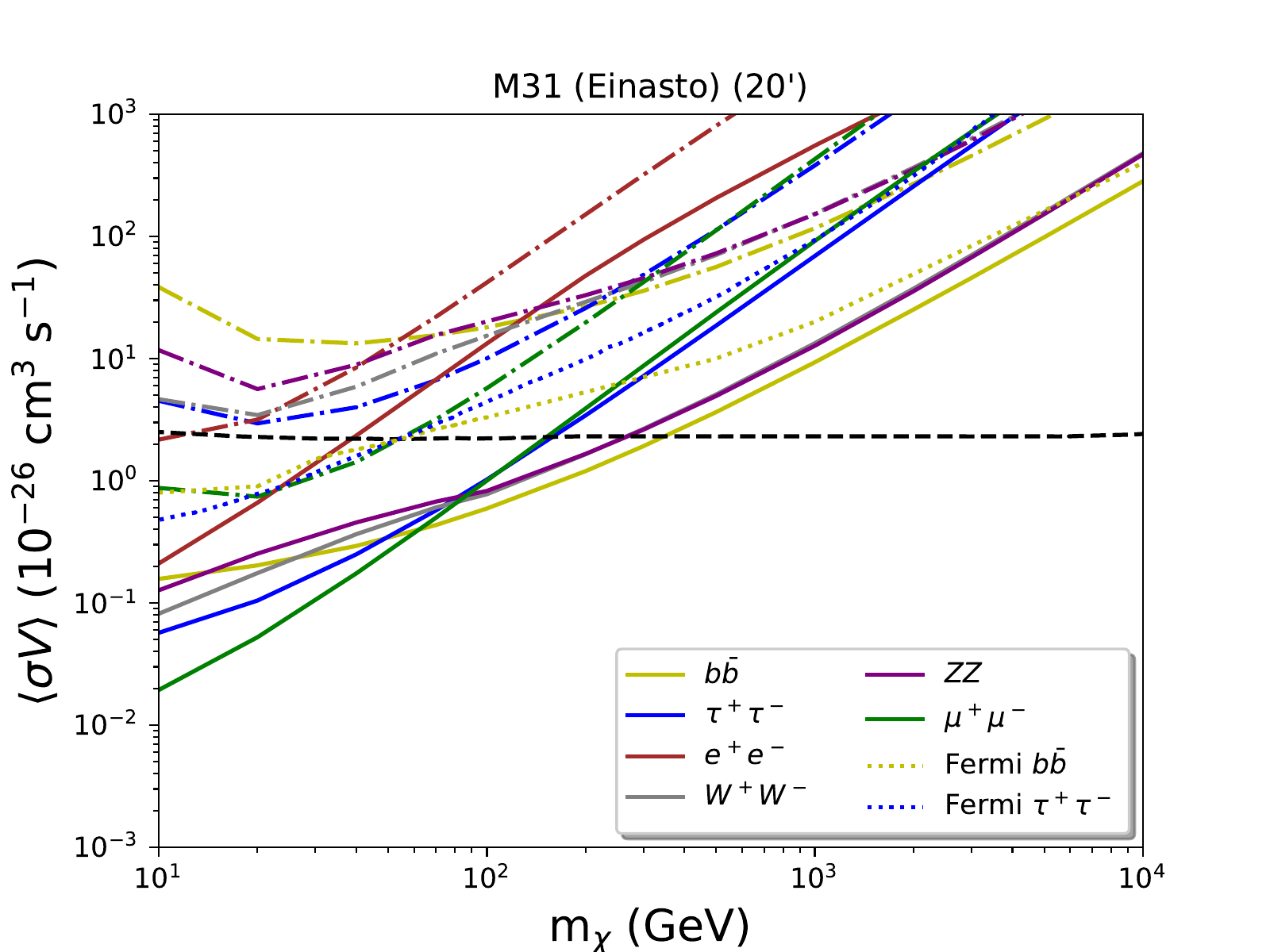}}
	\resizebox{0.49\hsize}{!}{\includegraphics{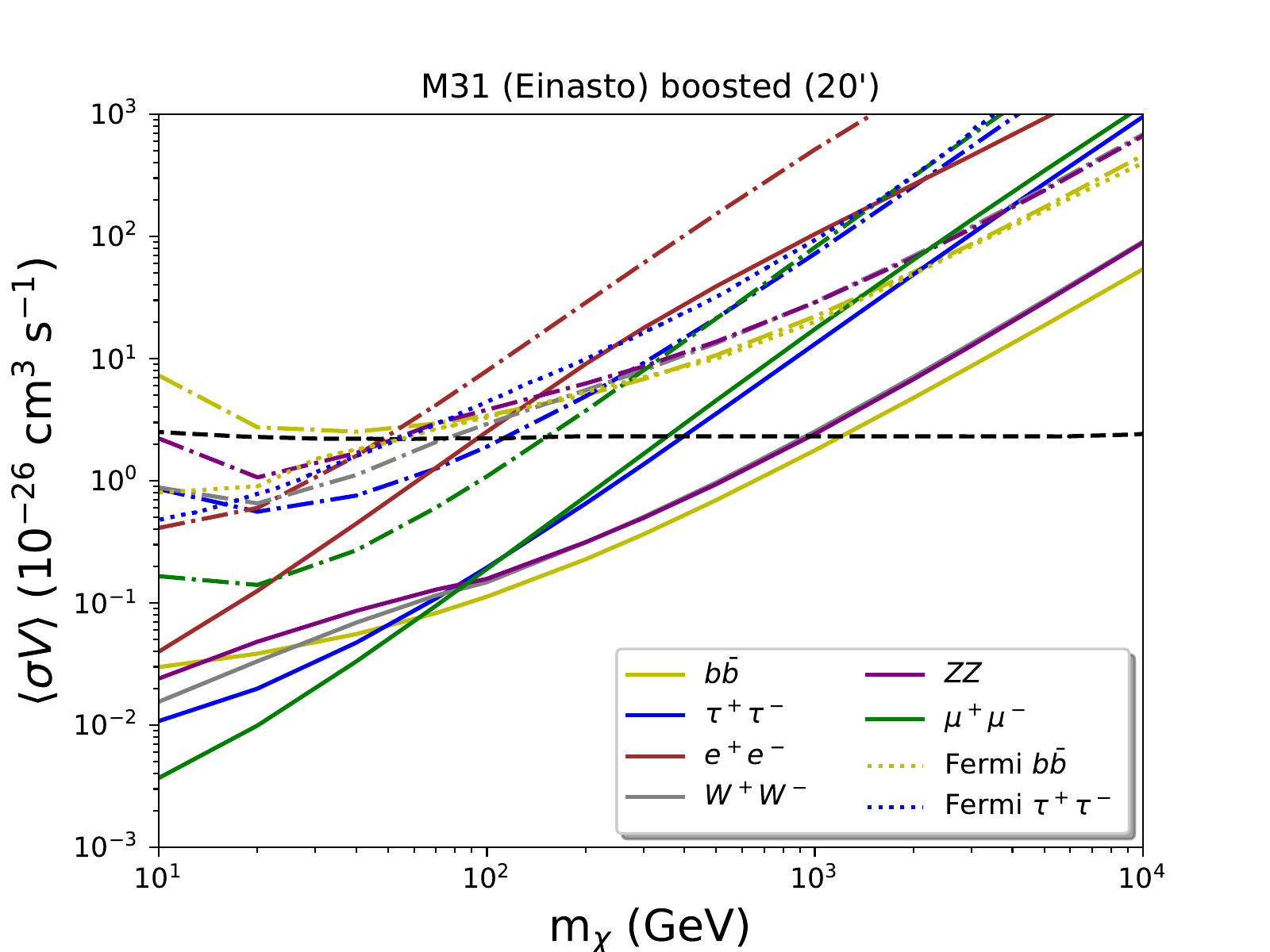}}
	\caption{M31 cross-section upper limits at $2\sigma$ confidence level for Einasto halos with a 20 arcminute ROI at 408 MHz. Left: unboosted. Right: boost factor $5.28$. The black dashed line shows the thermal relic cross-section~\cite{steigman2012}. The dotted lines show Fermi-LAT dwarf galaxy limits from \cite{Fermidwarves2016}. The solid lines show M31 results from this work for various annihilation channels using data points listed in Section~\ref{sec:data}, while dash-dotted lines do the same for the integrated flux from \cite{chan2019}.}
	\label{fig:m31-ein-med}
\end{figure}

In the case of a 20 arcminute ROI (or $\approx 4.5$ kpc) for the data from \cite{ficarra1985} (displayed in figs~\ref{fig:m31-nfw-med}, \ref{fig:m31-bur-med}, and \ref{fig:m31-ein-med}) the constraints are very similar to those found for largest 50 arcminute ROI, differing by around a factor of $\sim 1.5$ at all WIMP masses. At this choice of ROI the limits are still dominated by the data point from \cite{ficarra1985}, again demonstrating the power of low frequency data to constrain DM radio emissions. This case displays NFW results that are very similar at all displayed WIMP masses to the joint constraints from an exponential magnetic field model and $B_0 = 50$ $\mu$G used in \cite{egorov2013}.


\begin{figure}[htbp]
	\centering
	\resizebox{0.49\hsize}{!}{\includegraphics{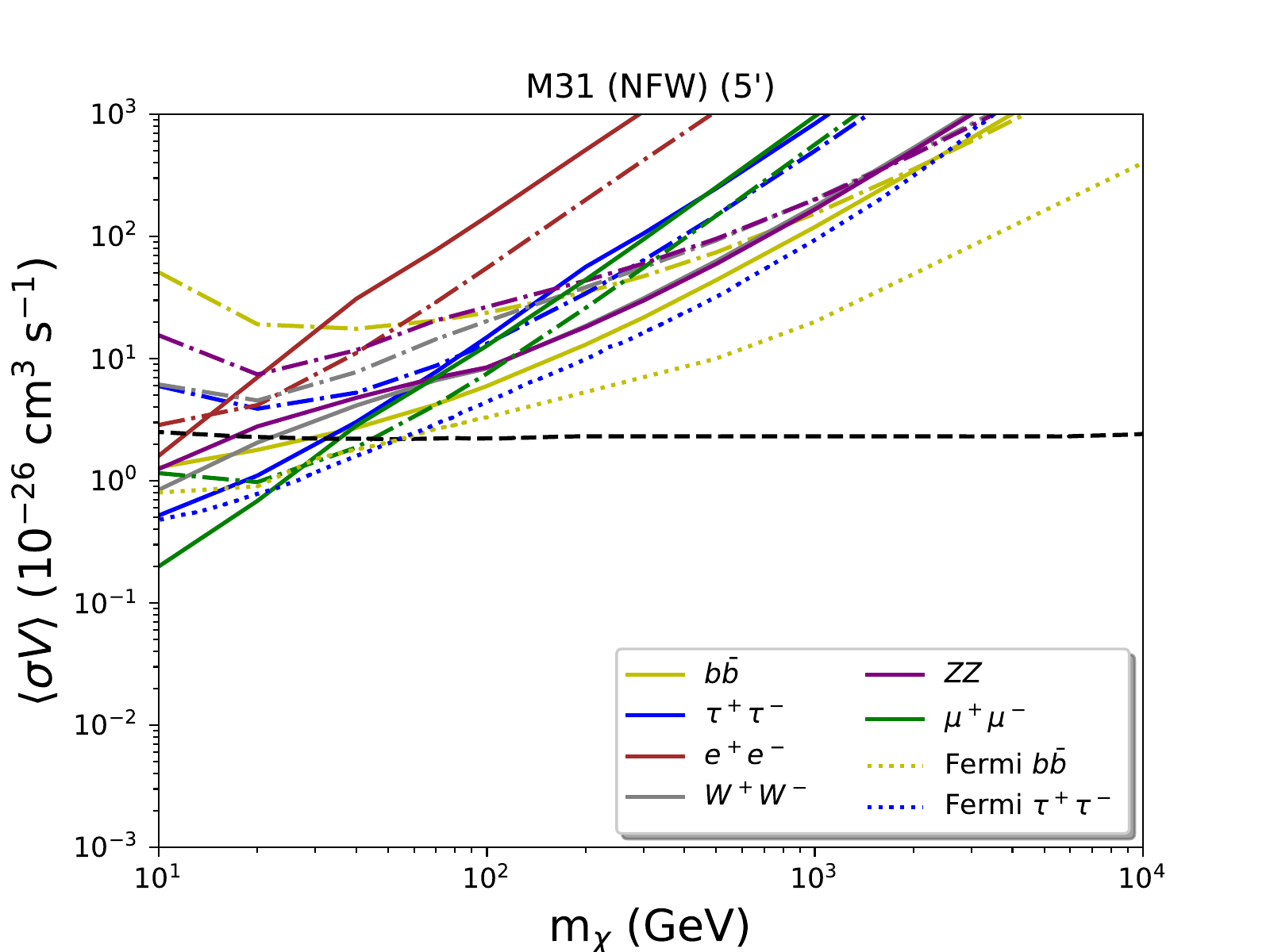}}
	\resizebox{0.49\hsize}{!}{\includegraphics{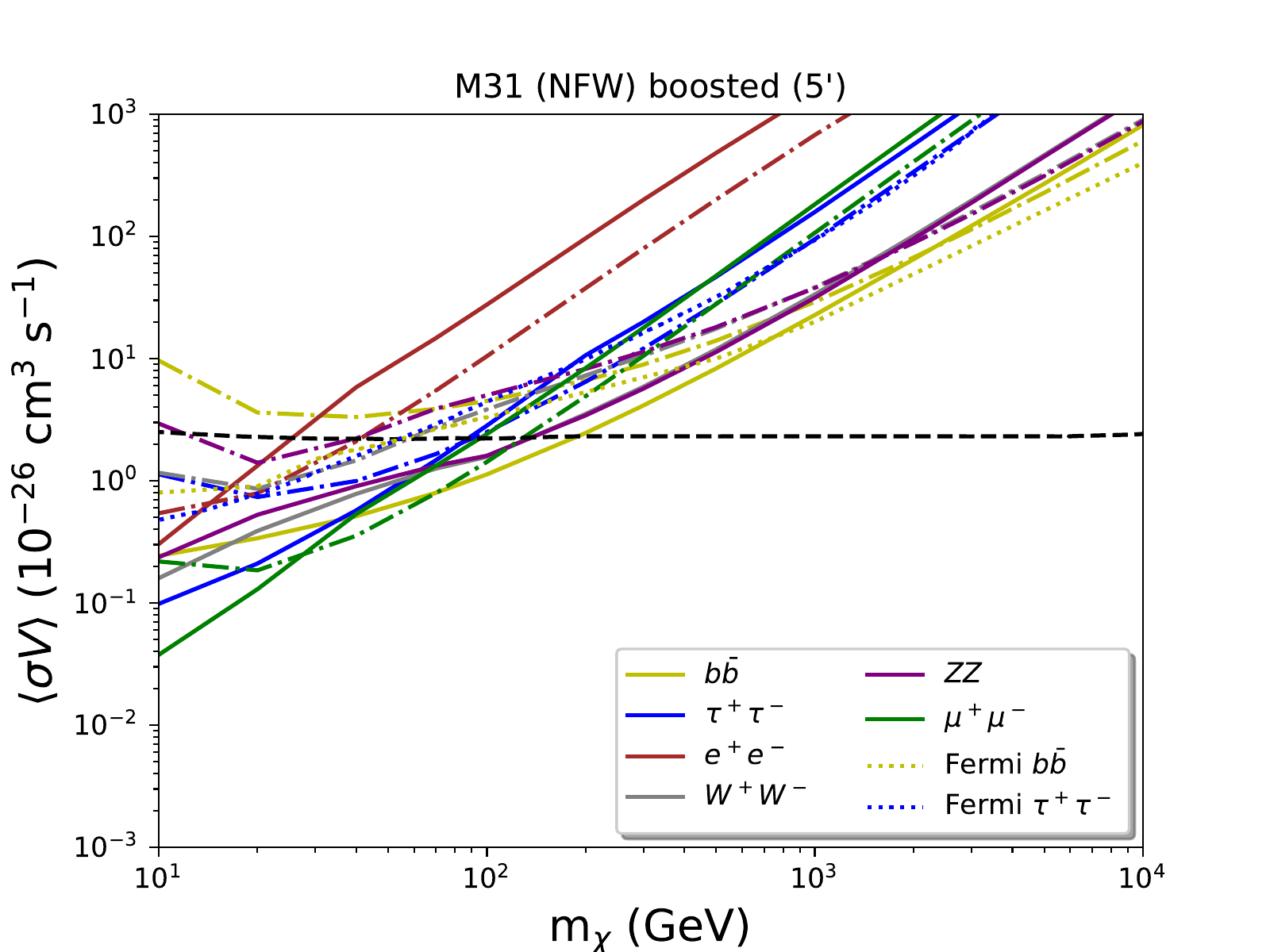}}
	\caption{M31 cross-section upper limits at $2\sigma$ confidence level for NFW halos with a 5 arcminute ROI at 408 MHz. Left: unboosted. Right: boost factor $5.28$. The black dashed line shows the thermal relic cross-section~\cite{steigman2012}. The dotted lines show Fermi-LAT dwarf galaxy limits from \cite{Fermidwarves2016}. The solid lines show M31 results from this work for various annihilation channels using data points listed in Section~\ref{sec:data}, while dash-dotted lines do the same for the integrated flux from \cite{chan2019}.}
	\label{fig:m31-nfw-pes}
\end{figure}

\begin{figure}[htbp]
	\centering
	\resizebox{0.49\hsize}{!}{\includegraphics{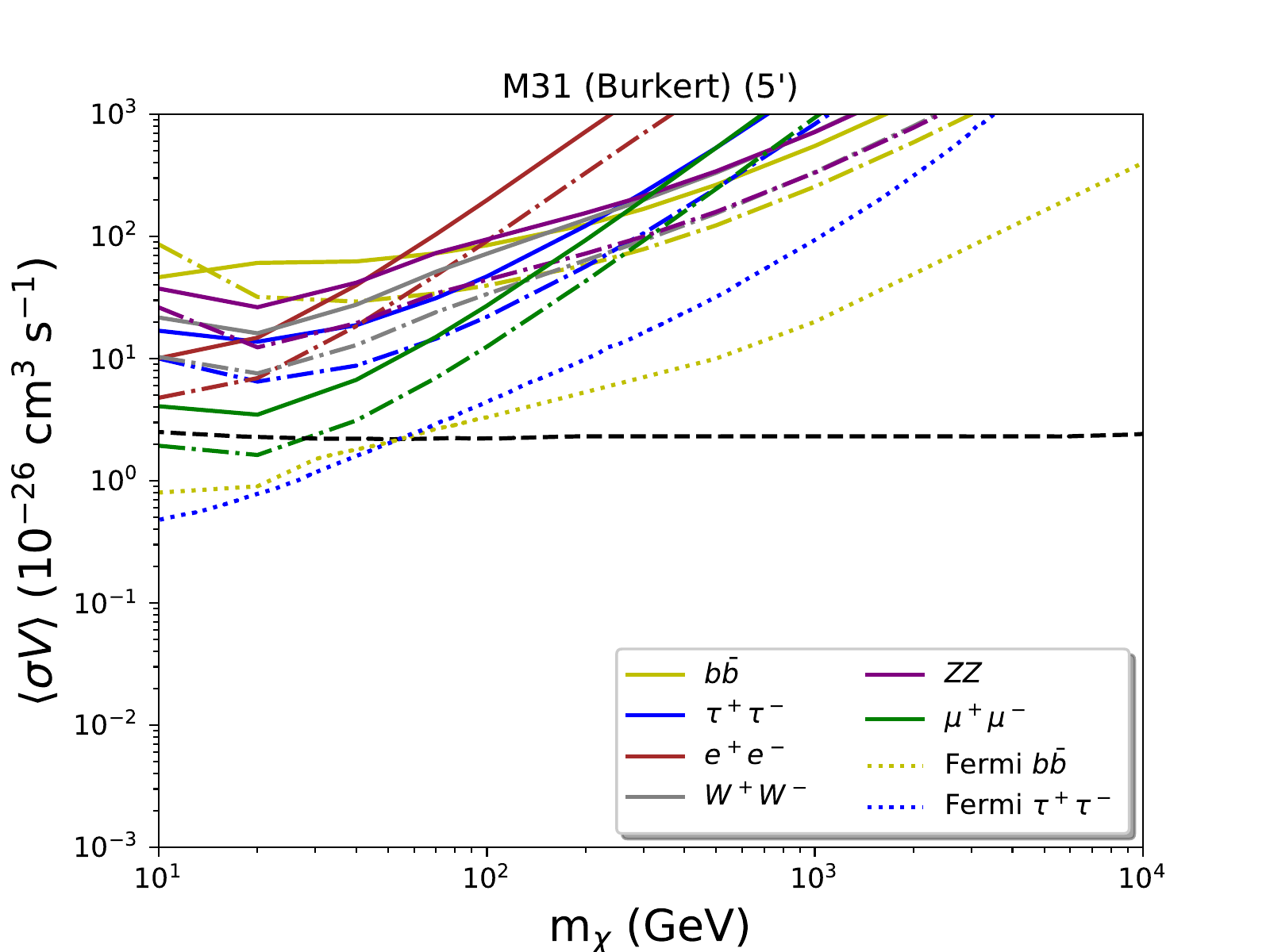}}
	\resizebox{0.49\hsize}{!}{\includegraphics{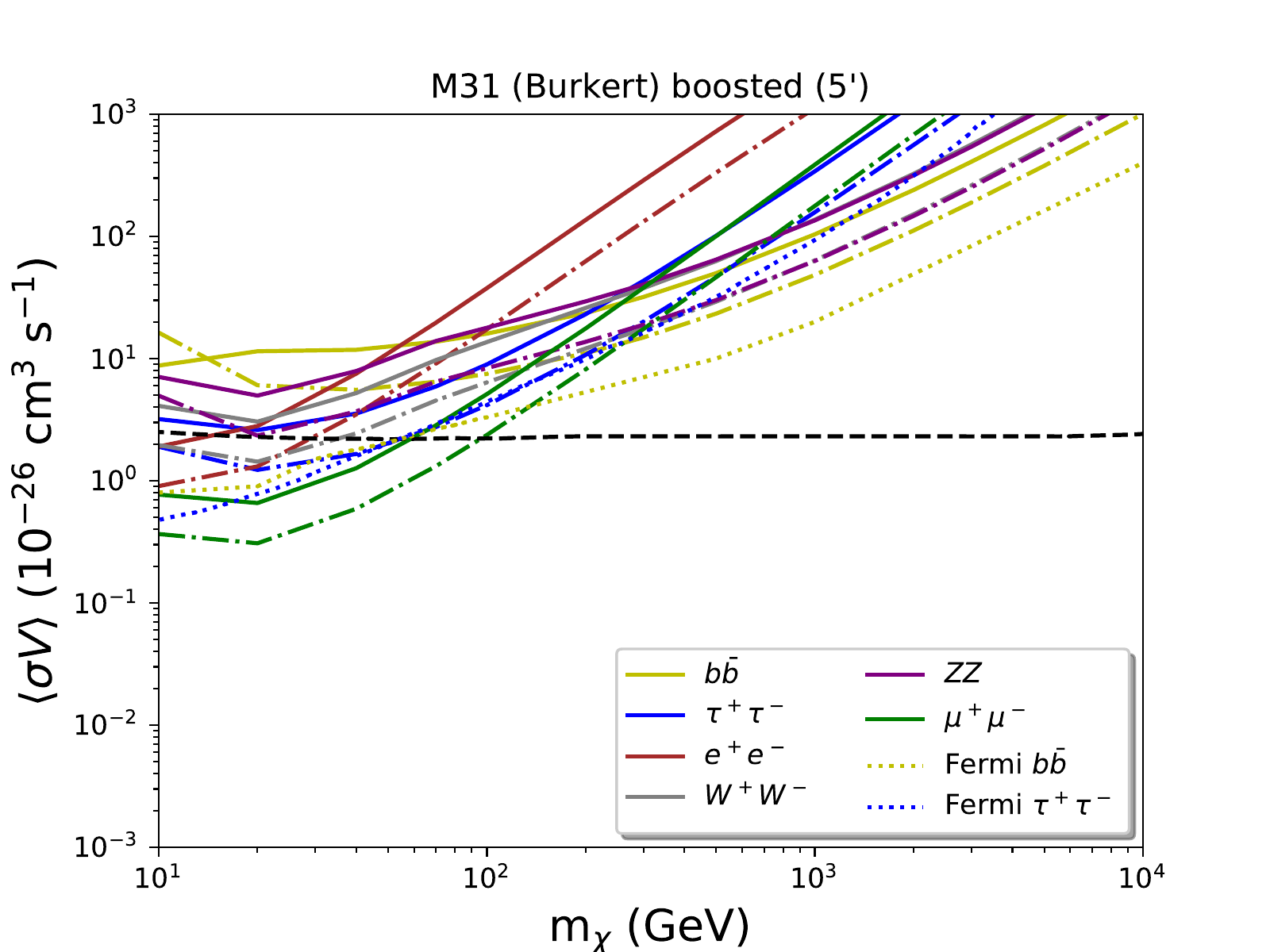}}
	\caption{M31 cross-section upper limits at $2\sigma$ confidence level for Burkert halos with a 5 arcminute ROI at 408 MHz. Left: unboosted. Right: boost factor $5.28$. The black dashed line shows the thermal relic cross-section~\cite{steigman2012}. The dotted lines show Fermi-LAT dwarf galaxy limits from \cite{Fermidwarves2016}. The solid lines show M31 results from this work for various annihilation channels using data points listed in Section~\ref{sec:data}, while dash-dotted lines do the same for the integrated flux from \cite{chan2019}.}
	\label{fig:m31-bur-pes}
\end{figure}

\begin{figure}[htbp]
	\centering
	\resizebox{0.49\hsize}{!}{\includegraphics{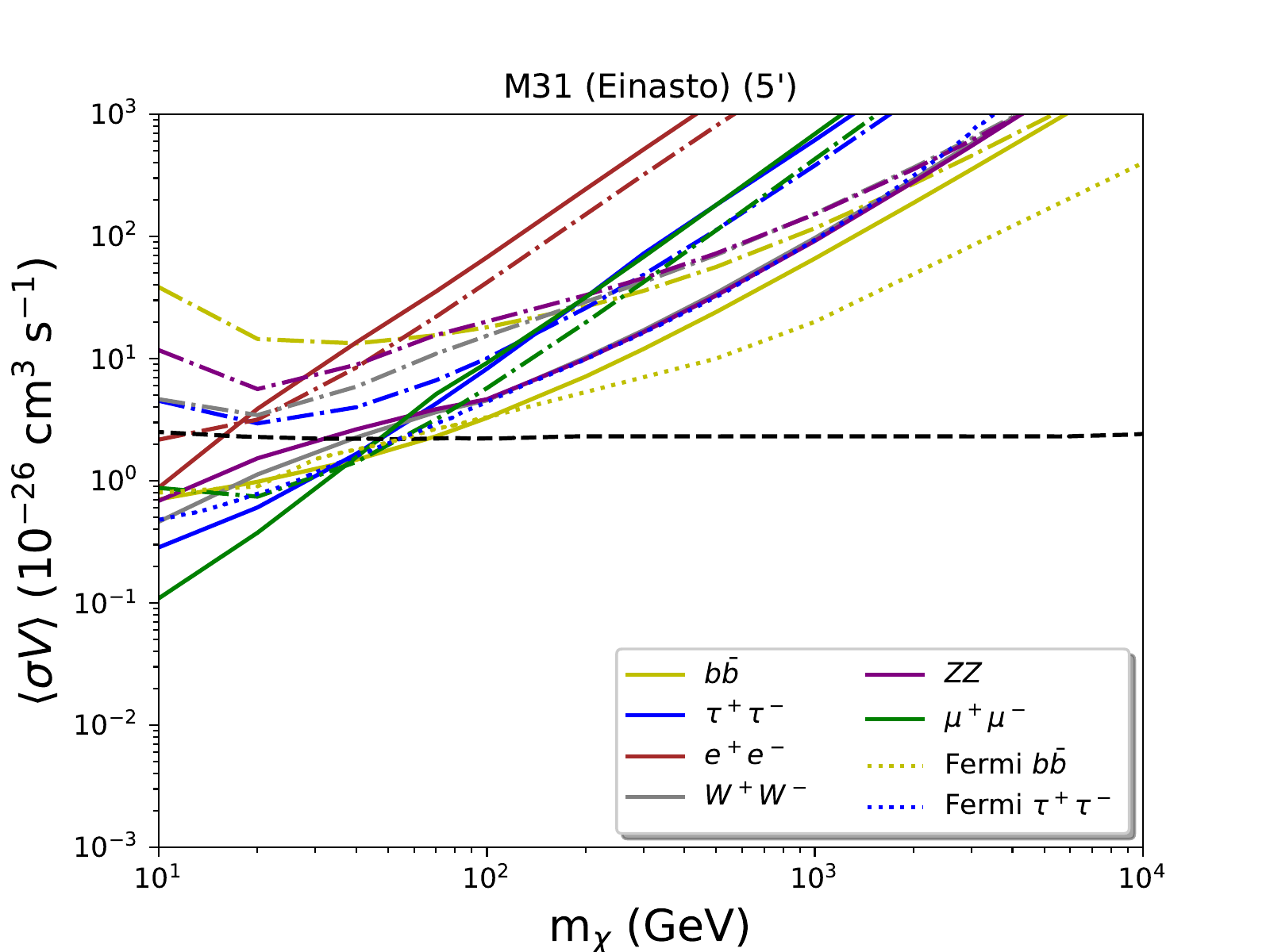}}
	\resizebox{0.49\hsize}{!}{\includegraphics{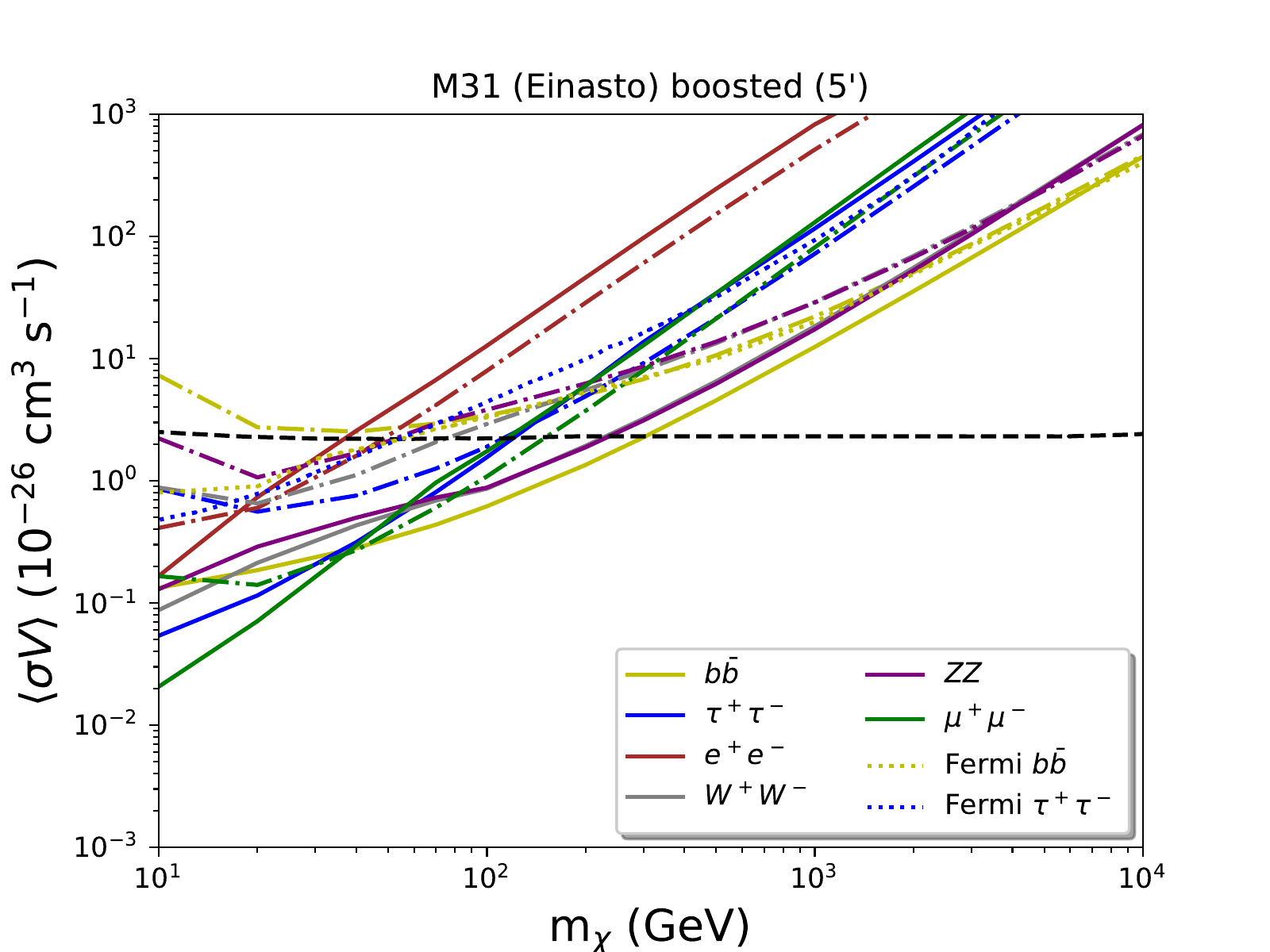}}
	\caption{M31 cross-section upper limits at $2\sigma$ confidence level for Einasto halos with a 5 arcminute ROI at 408 MHz. Left: unboosted. Right: boost factor $5.28$. The black dashed line shows the thermal relic cross-section~\cite{steigman2012}. The dotted lines show Fermi-LAT dwarf galaxy limits from \cite{Fermidwarves2016}. The solid lines show M31 results from this work for various annihilation channels using data points listed in Section~\ref{sec:data}, while dash-dotted lines do the same for the integrated flux from \cite{chan2019}.}
	\label{fig:m31-ein-pes}
\end{figure}

These results should be contrasted with Figs.~\ref{fig:m31-nfw-pes}, \ref{fig:m31-bur-pes}, \ref{fig:m31-ein-pes} which show the pessimistic ROI ($\sim 1$ kpc) for the 408 MHz data point from \cite{ficarra1985}. In this case, we can only probe below the relic cross-section for NFW and Einasto halo profiles with $m_{\chi} \lesssim 30$ GeV or $\lesssim 100$ GeV with the use of the boosting factor. Both of these halo profiles, with boosts, produced results that are slightly stronger than Fermi-LAT in the compared channels. With or without the boosting factor, the NFW and Einasto halos are competitive with the bounds from \cite{chan2019} data, and exceed them at low masses. In the Burkert case, the results are slightly weaker across the studied mass range than those obtained from \cite{chan2019} data. The NFW results for this pessimistic case agree quite closely with the joint constraints presented in \cite{egorov2013} for the most pessimistic magnetic field estimate both the $b$ quark and $\tau$ lepton channels (other channels are not displayed similarly in \cite{egorov2013}). We note that now the limits are dominated by the data points from \cite{ned_m31_jarrett2003,ned_m31_rice1988,ned_m31_still2009} at all masses and in all channels for the Burkert halo profile. In contrast, this is only true at masses above $70$ GeV for NFW and Einasto halos in the muon and electron channels, as well as the same halo profiles in $\tau$ for $m_\chi > 200$ GeV. However, in the case of $b$ quarks, $Z$, and $W$ channels the data from \cite{ficarra1985} still provides the dominant limit across the entire studied mass range in the NFW and Einasto cases. 


Importantly, we note that the ROI's used at 408 MHz are all substantially smaller than that used in the data of \cite{chan2019}. For the other data points from Section~\ref{sec:data} we used a 40 kpc ROI matching \cite{chan2019}. We justify this as they are taken at a similar frequency range and \cite{chan2019} notes a lack of other sources within this neighbourhood of the centre of M31 in their maps. In addition to this these fluxes are substantially larger, and so the impact of radial choice on our conclusions is very weak. We note that even had we used a 40 kpc ROI for the 408 MHz data point this would imply a $-0.4$ power-law index between the diffuse radio fluxes from \cite{chan2019} and \cite{ficarra1985}. The fact that we use substantially smaller radii is indicative then of the robustness of our results. We note that even in the case of a 40 kpc ROI our magnetic field is not necessarily unreliable, as the data used in the modelling of \cite{ruiz-granados2010} extends out to 38 kpc from the centre of M31.   

\section{Constraints from M33}
\label{sec:m33}
When studying M33 we use data points from \cite{chan2017} to compare to DM fluxes predicted within a radius of 7.5 kpc of the halo centre. The data point \cite{nvss1998} is compared to an ROI of 1.5 arcmin only. 

\begin{figure}[htbp]
	\centering
	\resizebox{0.49\hsize}{!}{\includegraphics{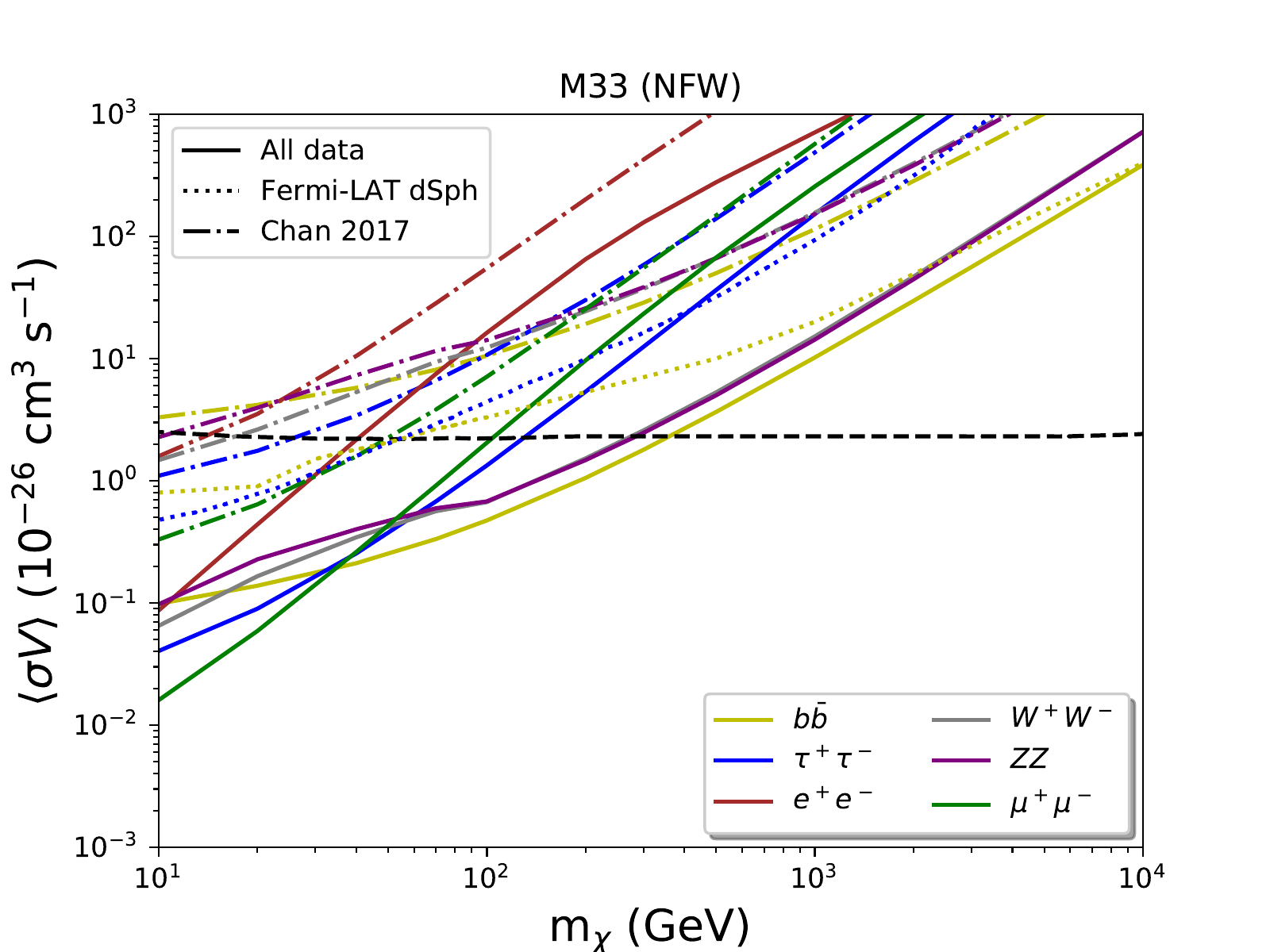}}
	\resizebox{0.49\hsize}{!}{\includegraphics{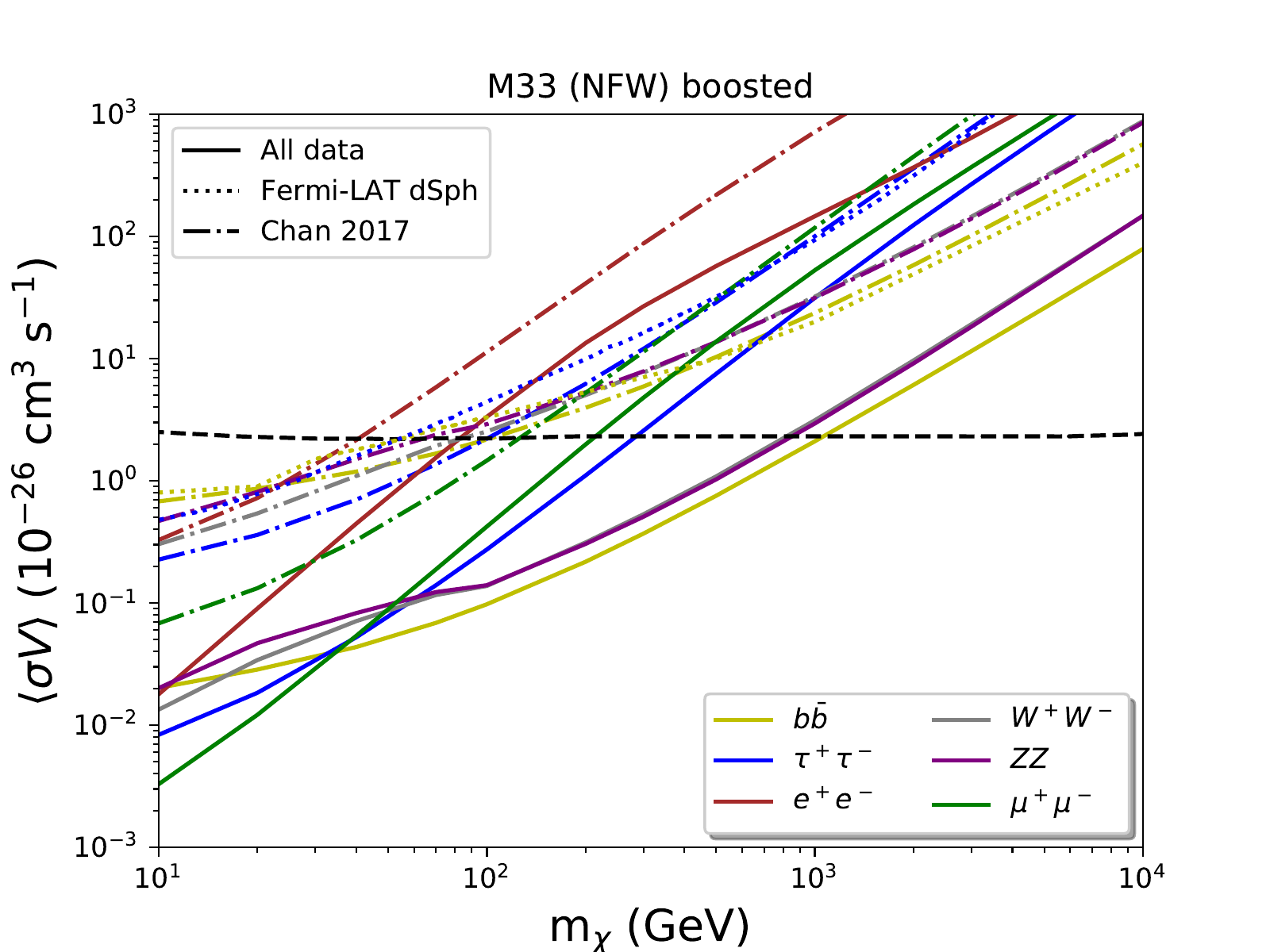}}
	\caption{M33 cross-section upper limits at $2\sigma$ confidence level for NFW halos. Left: unboosted. Right: boost factor $4.86$. The black dashed line shows the thermal relic cross-section~\cite{steigman2012}. The dotted lines show Fermi-LAT dwarf galaxy limits from \cite{Fermidwarves2016}. The solid lines show M33 results from this work for various annihilation channels. The dot-dashed lines show our calculations using only the data points from \cite{chan2017}.}
	\label{fig:m33-nfw}
\end{figure}

\begin{figure}[htbp]
	\centering
	\resizebox{0.49\hsize}{!}{\includegraphics{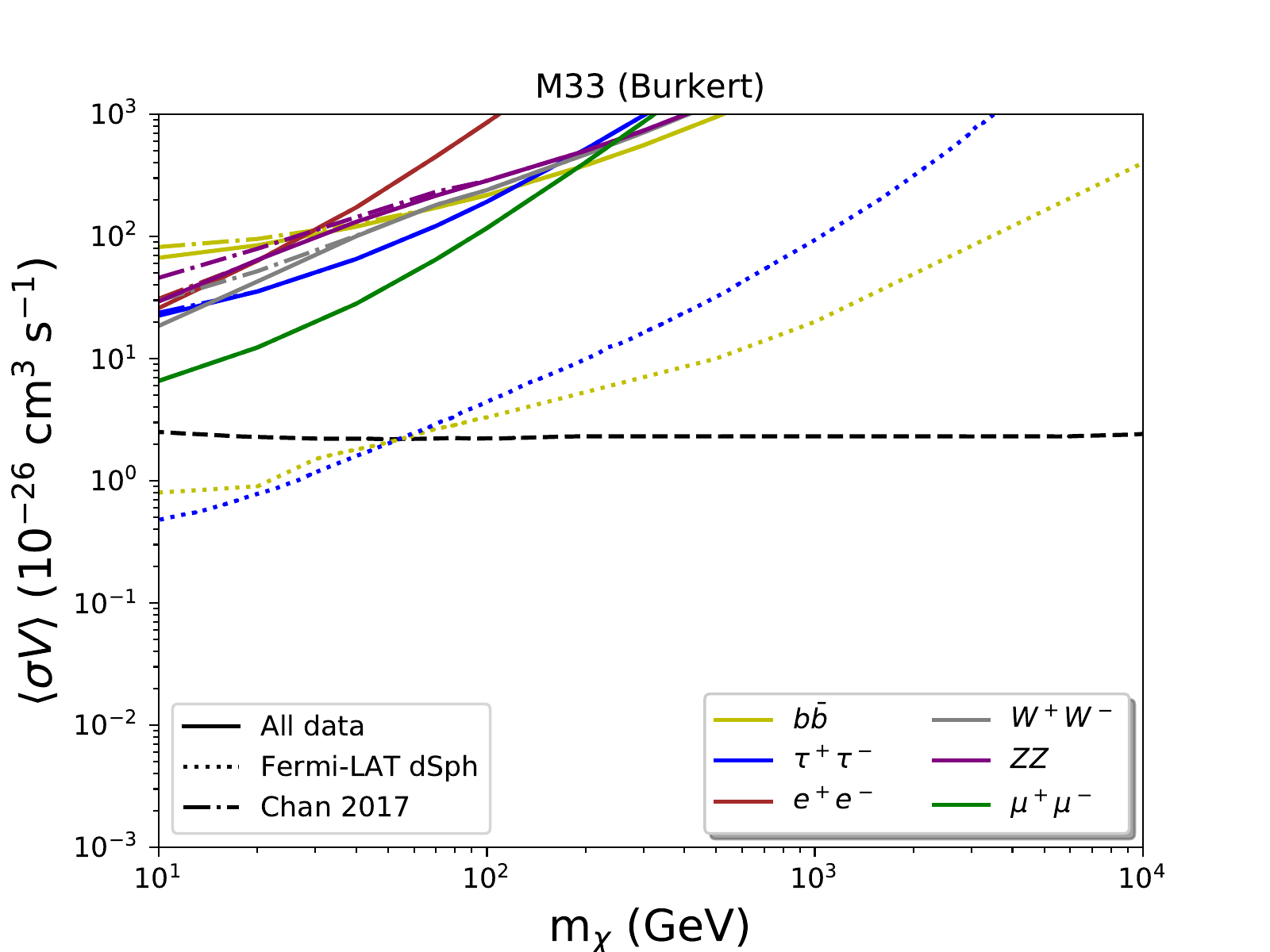}}
	\resizebox{0.49\hsize}{!}{\includegraphics{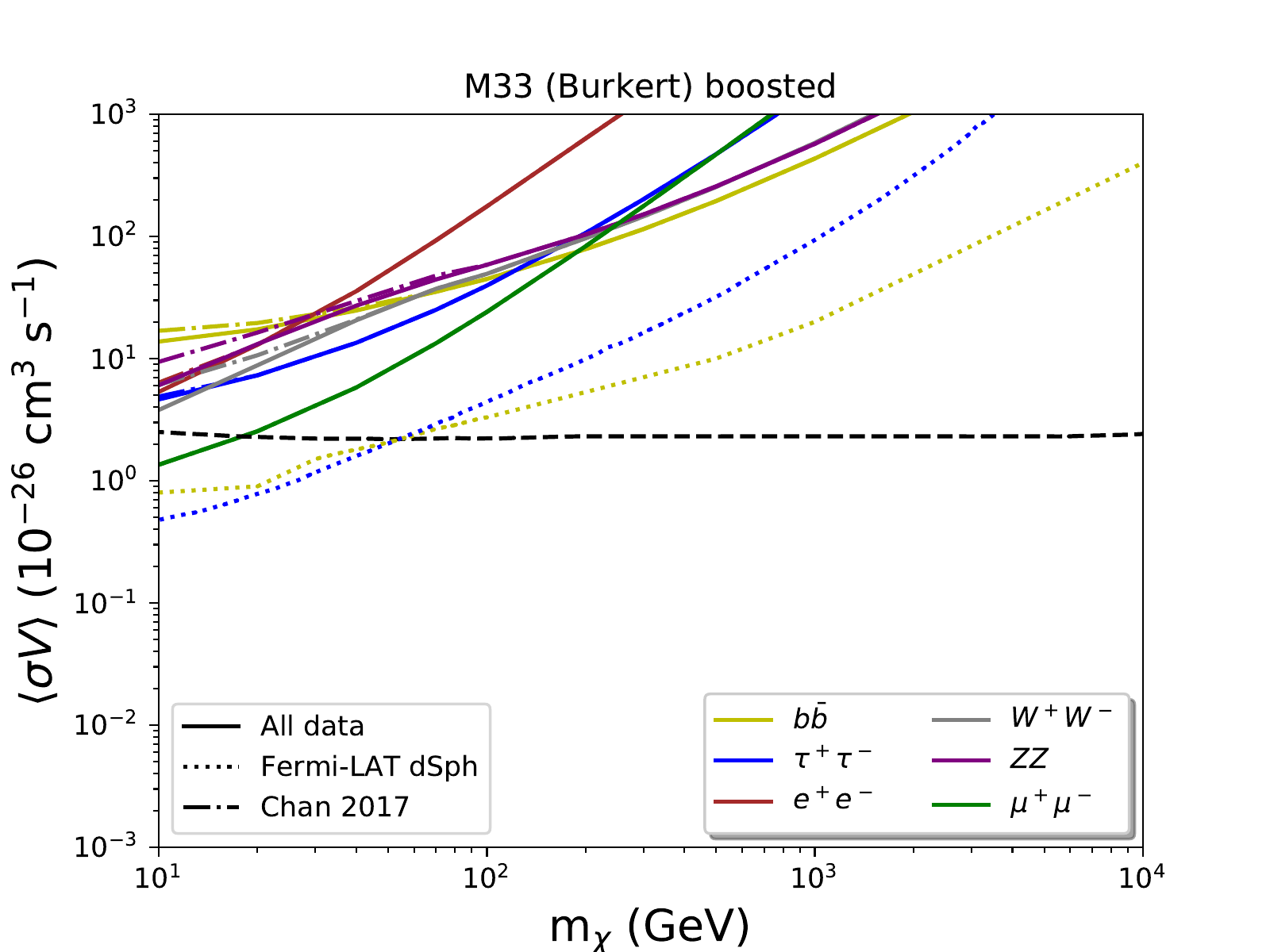}}
	\caption{M33 cross-section upper limits at $2\sigma$ confidence level for Burkert halos. Left: unboosted. Right: boost factor $4.86$. The black dashed line shows the thermal relic cross-section~\cite{steigman2012}. The dotted lines show Fermi-LAT dwarf galaxy limits from \cite{Fermidwarves2016}. The solid lines show M33 results from this work for various annihilation channels. The dot-dashed lines show our calculations using only the data points from \cite{chan2017}.}
	\label{fig:m33-bur}
\end{figure}

In Fig.~\ref{fig:m33-nfw} we display cross-section constraints found using an NFW density profile and the radio limits presented in \cite{chan2017} as well as \cite{nvss1998}. Both boosted (left) and unboosted (right) cases are displayed, in the former case using a boost factor of $4.86$ following~\cite{chan2017}. Despite a strong central field value of $16.51$ $\mu$G, the \cite{chan2017}-only constraints for this case are only competitive with Fermi-LAT when when a boosting factor is employed. However, when the data from \cite{nvss1998} is included the resulting limits are better than those from Fermi-LAT WIMP masses $<100$ GeV by around an order of magnitude, becoming similar at larger masses. This also allows us to probe below the relic cross-section value up to around $100$ GeV WIMP mass in all channels displayed. It is noteworthy that results using only data from \cite{chan2017} presented here agree quite closely those found in \cite{chan2017}, being slightly stronger at low WIMP masses. However, the limits presented here are at least an order of magnitude weaker for the electron-positron annihilation channel. The difference between this work and \cite{chan2017} is our use of the radially varying field (with average values consistent with \cite{chan2017}), consideration of all energy-loss processes, and the inclusion of diffusion. This demonstrates that the seemingly reasonable assumptions made in \cite{chan2017}, that the magnetic field is constant and diffusion is negligible, can result in an over-estimation of the predicted radio flux for some annihilation channels. However, the assumptions of \cite{chan2017} do seem to provide conservative limits for other channels, in agreement with ours for an exponential field and diffusion. It is also important to note that the limits depend very strongly on the assumed halo profile as the Burkert case in Fig.~\ref{fig:m33-bur} is at least 1(2) orders of magnitude less constraining than for an NFW density profile with \cite{chan2017}(\cite{nvss1998}) data (the Burkert case being more sensitive to a steep field profile and diffusion). Interestingly, the Burkert limits are in agreement for both data sets used.

This has important consequences for the conclusions drawn in \cite{chan2017}. Notably that their claimed tension with AMS-02 positron excess models from \cite{dimauro2015} weakens due to the weakening of constraints for WIMP masses above $10$ GeV with the electron annihilation channel. The muon channel can maintain a marginal tension only when a boost factor is assumed (tau is unchanged from \cite{chan2017}). Of course, these tensions disappear entirely when a Burkert density profile is used instead. However, it can be shown that the tension with models favoured by the Galactic Centre GeV excess~\cite{abazajian2016} remains, although other models such as \cite{calore2015} are unaffected by the M33 data. The impactfulness of the M33 data can be recovered somewhat when using \cite{nvss1998} data, so that the claimed tensions from \cite{chan2017} are once again in force for muon and quark channels but only for an NFW halo profile.


\section{The Excesses revisited}
\label{sec:ex}

In this section we examine DM models proposed to explain the three excesses: AMS-02~\cite{cholis2019}, Galactic Centre GeV gamma-rays~\cite{calore2015}, and DAMPE~\cite{dampedm1}. 

\begin{figure}[htbp]
	\centering
	\resizebox{0.49\hsize}{!}{\includegraphics{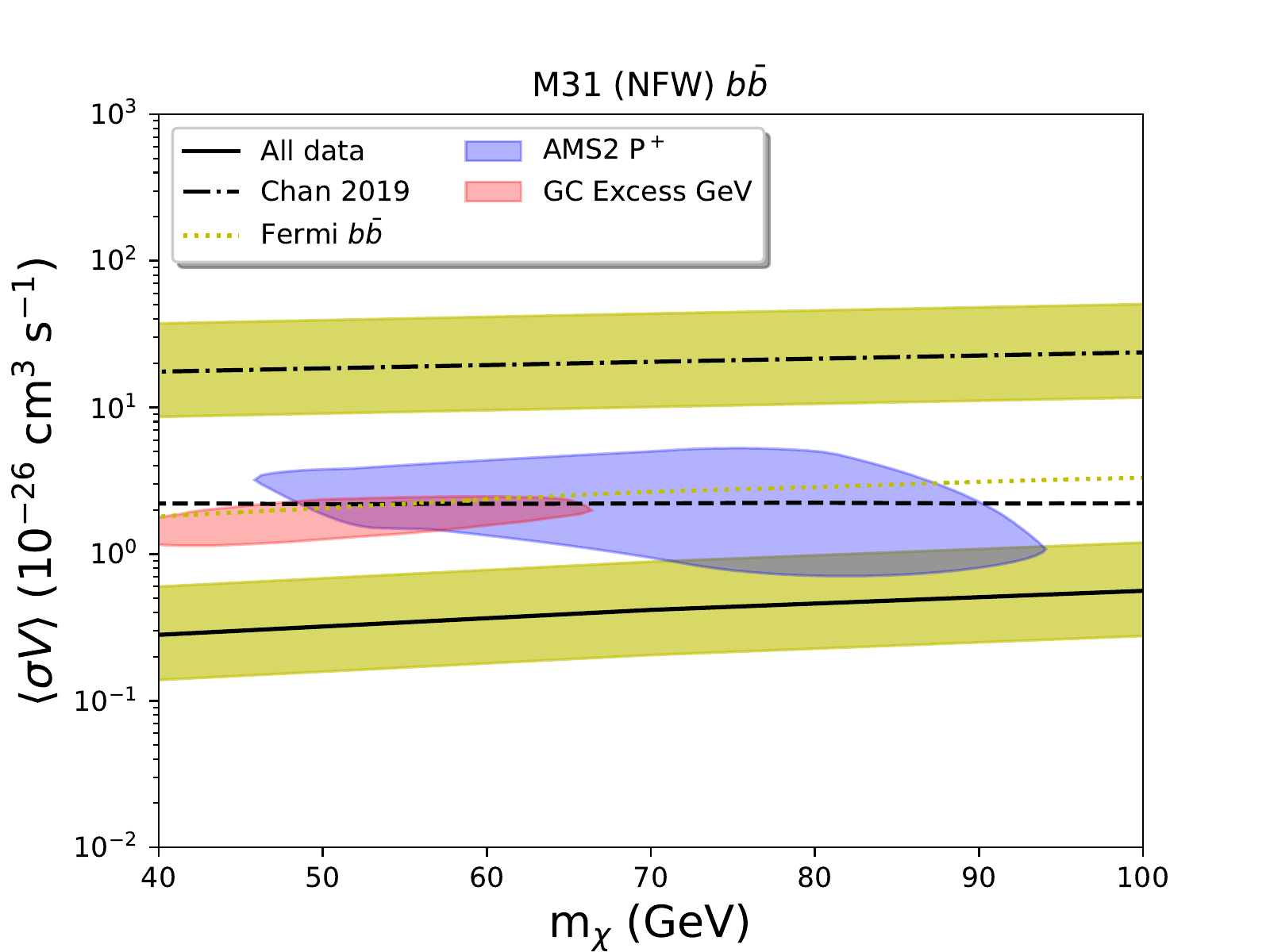}}
 	\resizebox{0.49\hsize}{!}{\includegraphics{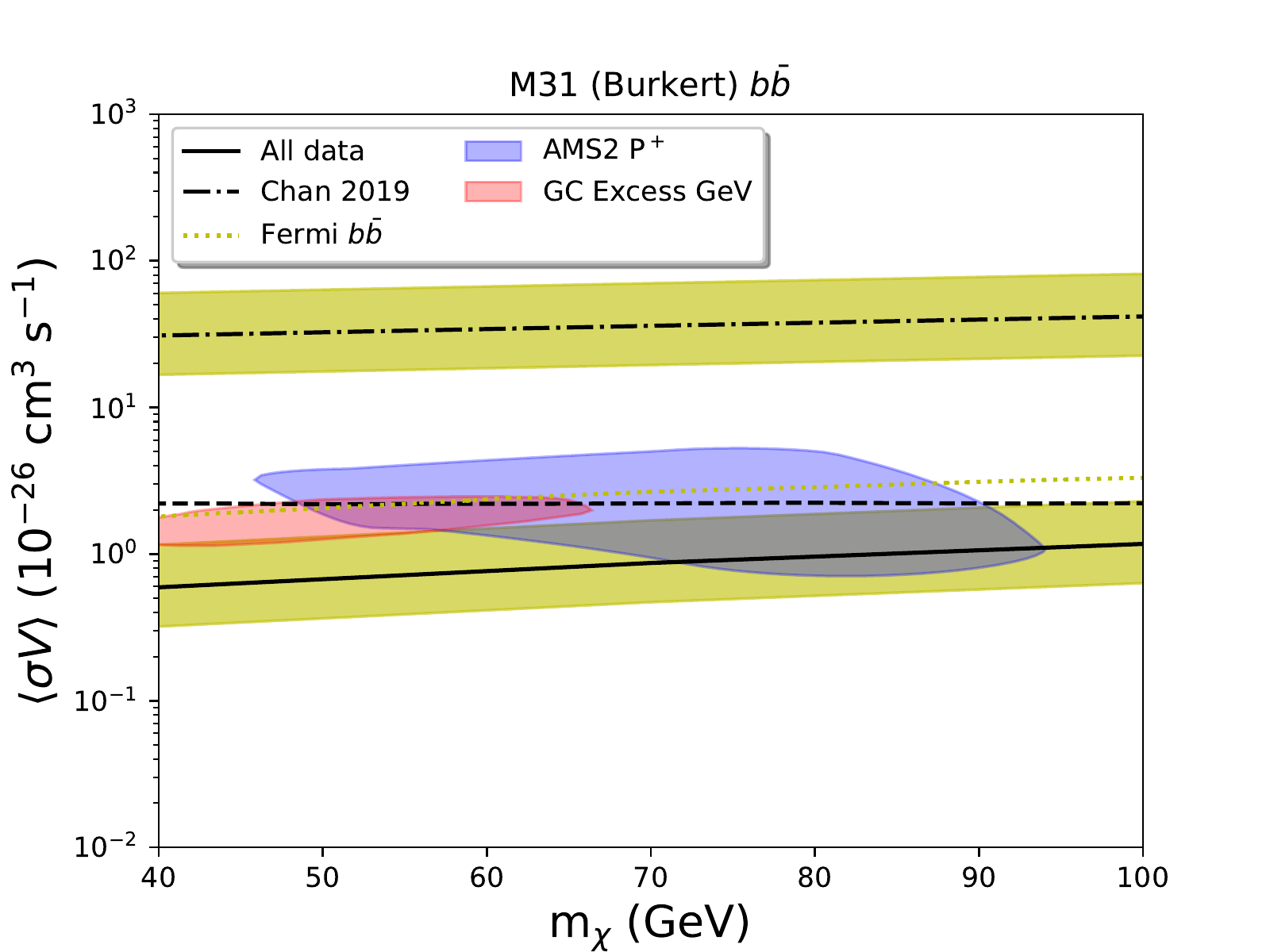}}
	\resizebox{0.49\hsize}{!}{\includegraphics{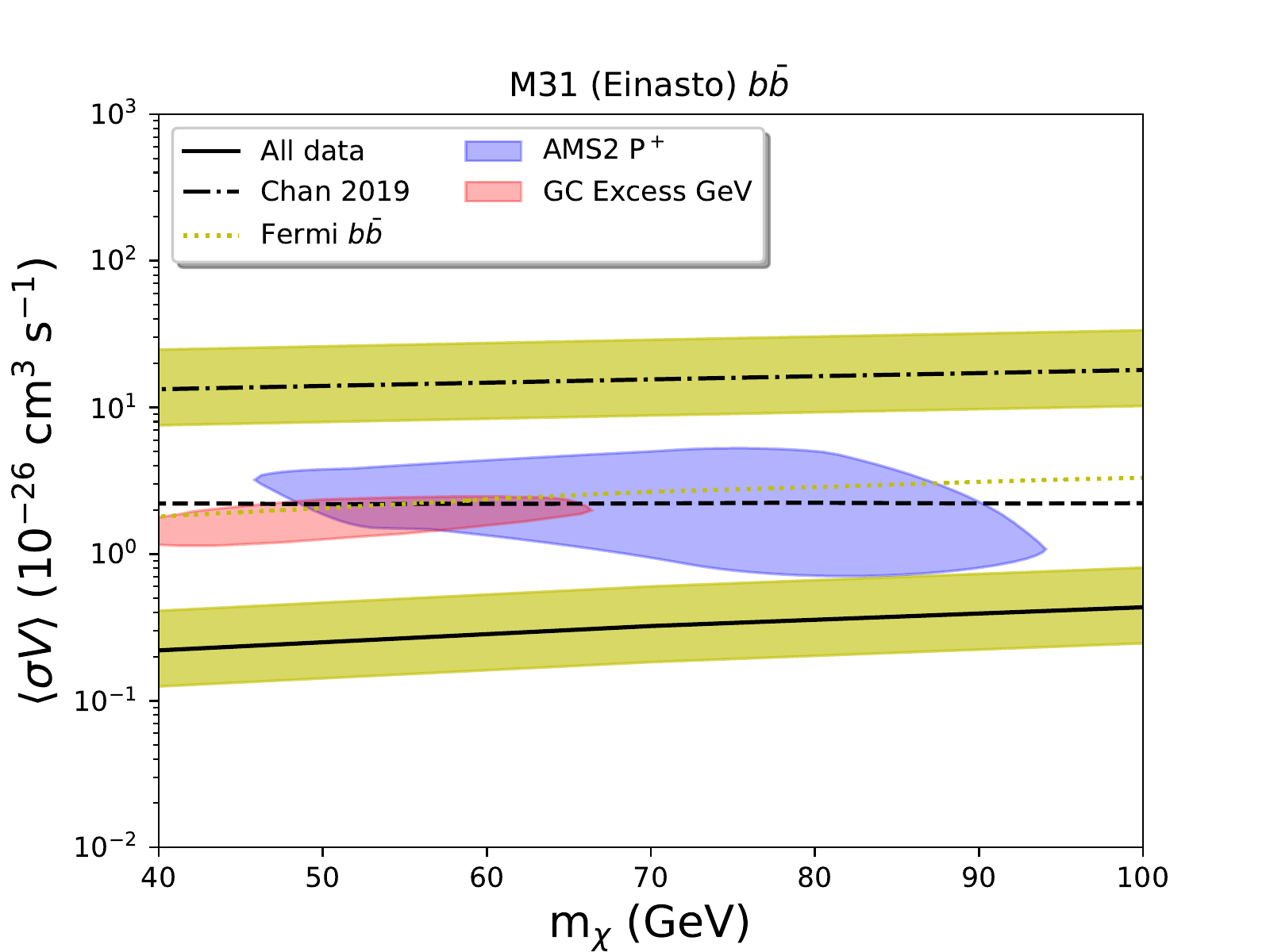}}
	\caption{M31 cross-section upper limits at $2\sigma$ confidence level for NFW (upper left), Burkert (upper right), and Einasto (lower) halos. Solid lines show limits from the spectra listed in Section~\ref{sec:data} while dash-dotted lines are for the integrated flux from \cite{chan2019}. Shading around the lines displays uncertainties. The blue shaded region shows the parameter space for AMS-02 anti-proton excess DM models~\cite{cholis2019} and the red represents the Galactic Centre GeV gamma-ray excess models from \cite{calore2015}. The black dashed line shows the thermal relic cross-section~\cite{steigman2012}.}
	\label{fig:m31-ex1}
\end{figure}

In Fig.~\ref{fig:m31-ex1} we compare constraints from derived M31 to the favoured parameter space regions for both the GeV gamma-rays and the AMS-02 excesses. We use both the integrated flux for M31 from \cite{chan2019} (`Chan 2019' in the plots) and data points quoted in Section~\ref{sec:data} (`All data' in plots). Uncertainties for the cross-sections are also displayed using those from $r_s$, $\rho_s$, and $B_0$. It is evident that for the integrated flux from \cite{chan2019} that only the Einasto case allows for potential constraint of the parameter space for AMS-02, however, this is very marginal. In contrast, the data point from \cite{ficarra1985} at 408 MHz yields constraints that can rule out the entire parameter space of both excesses regardless of the choice of halo density profile (for the displayed WIMP mass range). This is significant as the observations made in \cite{ficarra1985} had an angular resolution at the levl of a few arcminutes, and thus suited for probing large-scale diffuse emission in M31. In the pessimistic 5 arcminute ROI case for \cite{ficarra1985} we can rule out the parameter spaces only with the aid of a boosting factor in the NFW and Einasto halo profiles, the Burkert case provides no constraining power.
 
%
%

\begin{figure}[htbp]
	\centering
	\resizebox{0.49\hsize}{!}{\includegraphics{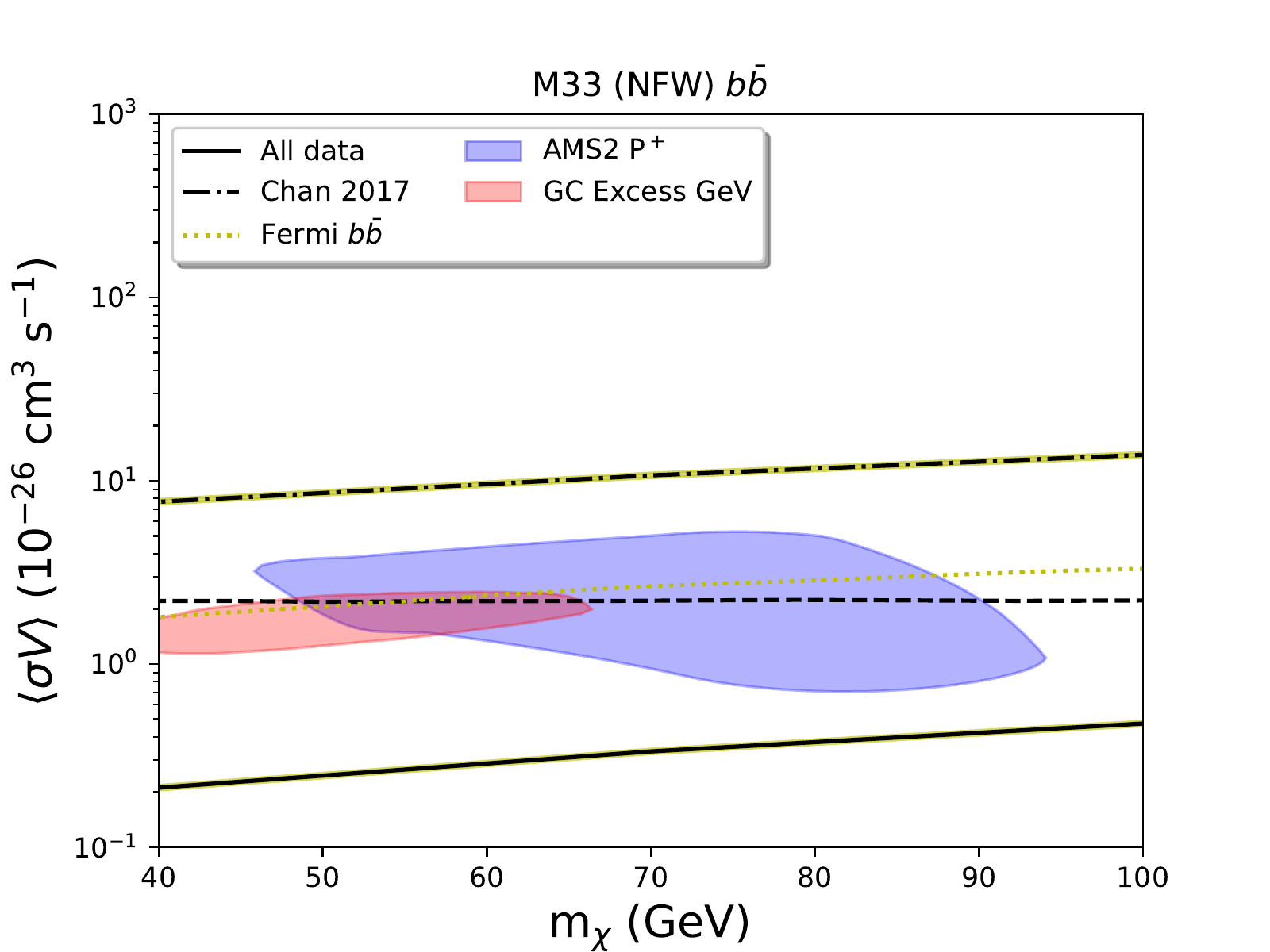}}
	\resizebox{0.49\hsize}{!}{\includegraphics{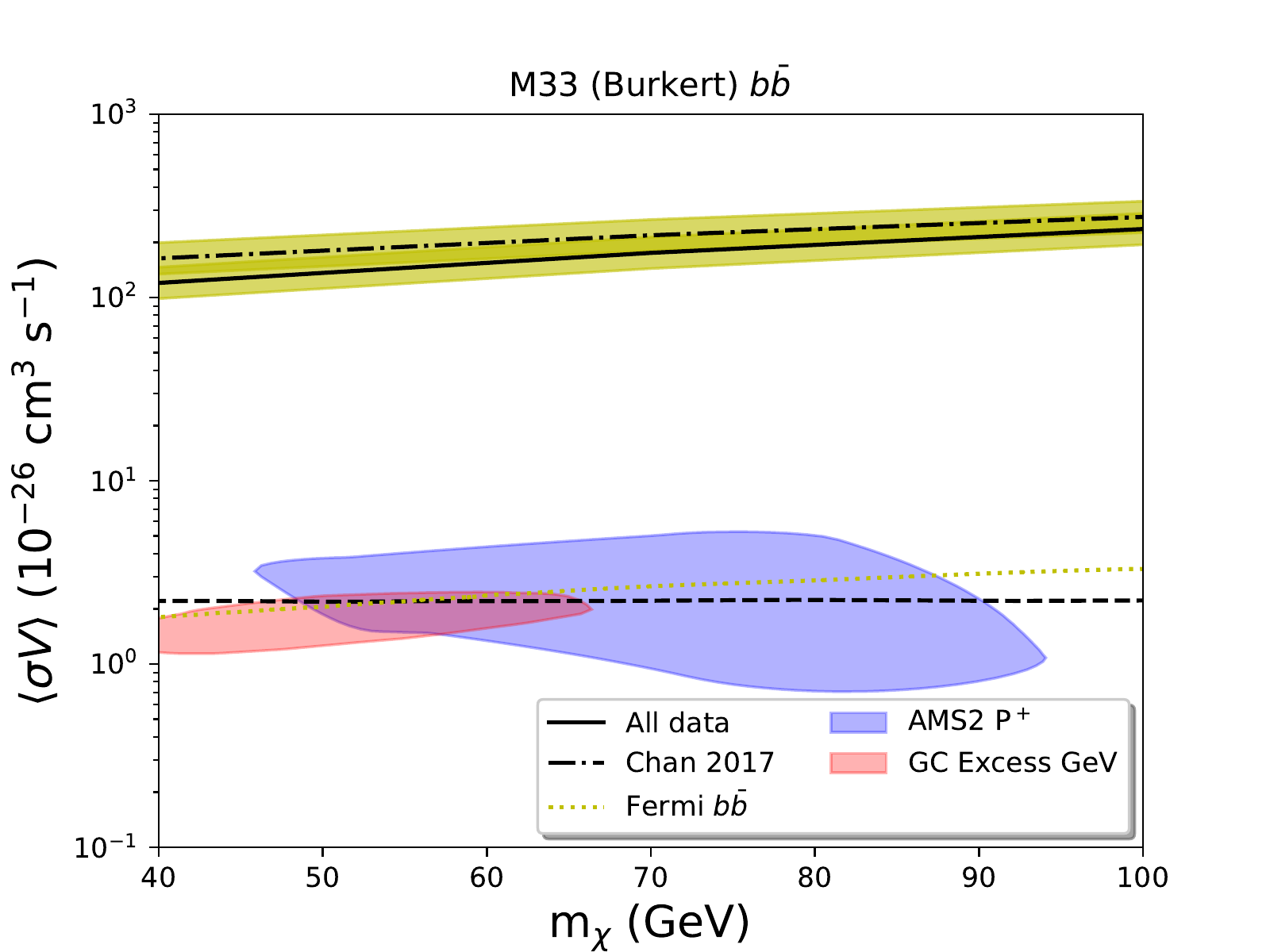}}
	\caption{M33 cross-section upper limits at $2\sigma$ confidence level for NFW (left) and Burkert (right) halos. Solid lines show limits from the spectra listed in Section~\ref{sec:data} while dash-dotted lines are from \cite{chan2017} data only. Shading around the lines displays uncertainties. The blue shaded region shows the parameter space for AMS-02 anti-proton excess DM models~\cite{cholis2019} and the red represents the Galactic Centre GeV gamma-ray excess models from \cite{calore2015}. The black dashed line shows the thermal relic cross-section~\cite{steigman2012}.}
	\label{fig:m33-ex1}
\end{figure}

Figure~\ref{fig:m33-ex1} displays results analogous to Fig.~\ref{fig:m31-ex1} but for M33. For both NFW (left) and Burkert (right) density profiles, using only data from \cite{chan2017} (dash-dotted line) does not allow for the constraint of either of the parameter spaces of interest. However, the inclusion of data from \cite{nvss1998} allows us to completely rule out all of the parameter space assigned by AMS-02 anti-proton and Galactic Centre gamma-ray excesses with an NFW density profile. However, the relevance of diffusion strongly affects the Burkert case.  

\begin{figure}[htbp]
	\centering
	\resizebox{0.49\hsize}{!}{\includegraphics{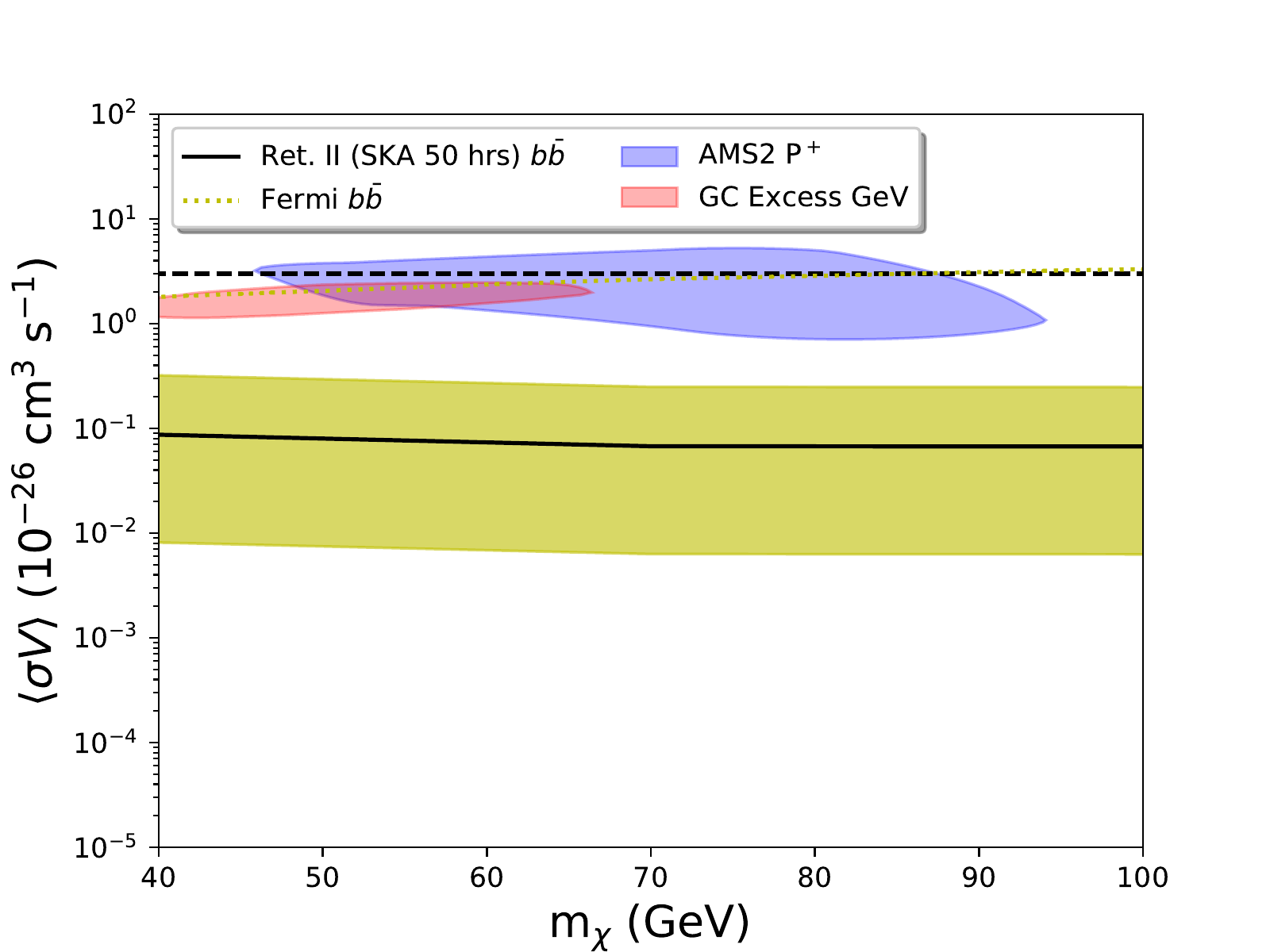}}
	\resizebox{0.49\hsize}{!}{\includegraphics{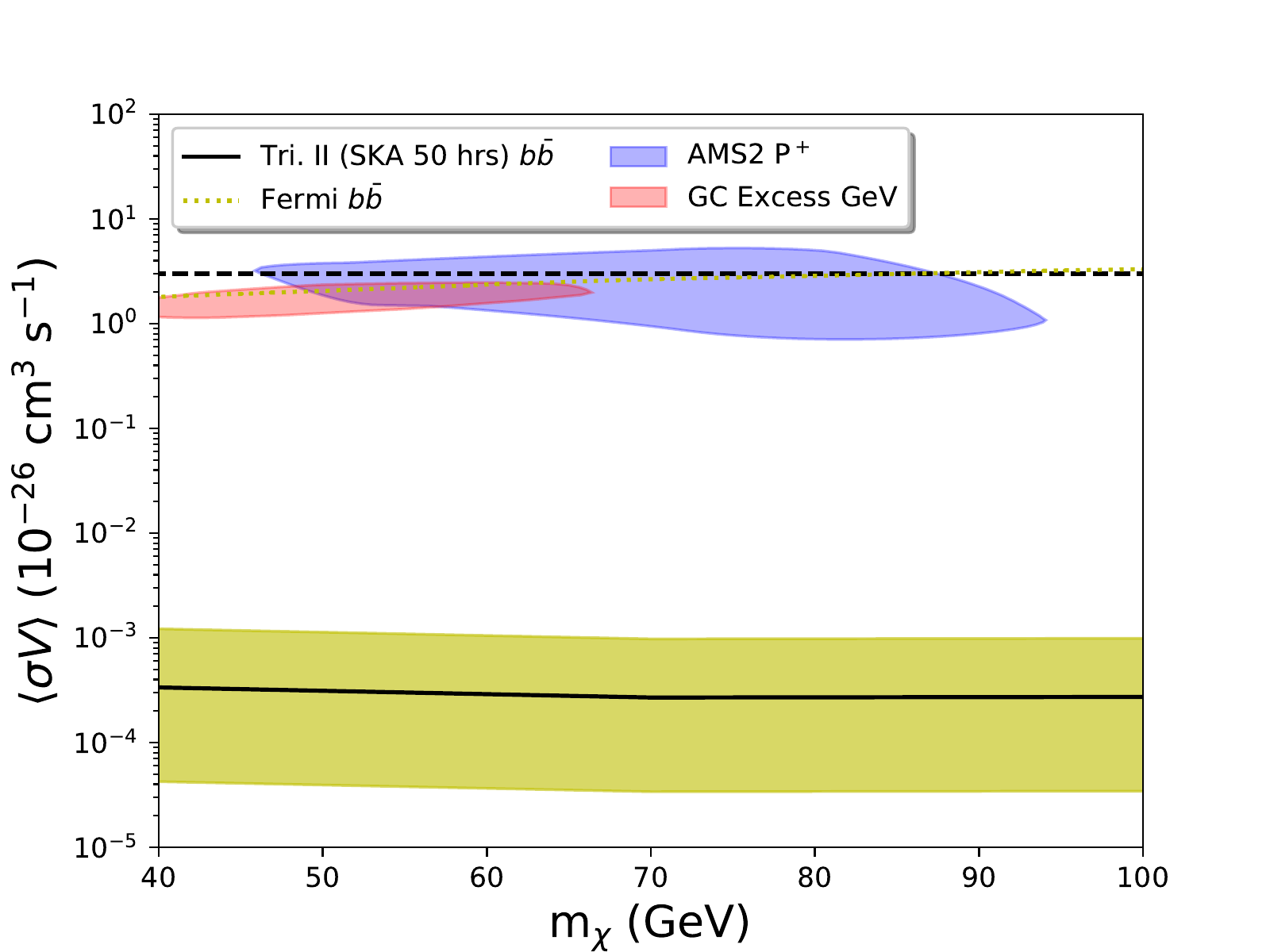}}
	\caption{Dwarf galaxy non-observation cross-section upper limits at $2\sigma$ confidence level with 0.5 degree ROI. Left: Reticulum II. Right: Triangulum II. Shading around the upper-limit lines displays uncertainties. The blue shaded region shows the parameter space for AMS-02 anti-proton excess DM models~\cite{cholis2019} and the red represents the Galactic Centre GeV gamma-ray excess models from \cite{calore2015}. The black dashed line shows the thermal relic cross-section~\cite{steigman2012}.}
	\label{fig:dsph-ex1}
\end{figure}

In Fig.~\ref{fig:dsph-ex1} we display SKA non-observation constraints at $2\sigma$ confidence level for two chosen dwarf spheroidal galaxies, these being Reticulum II and Triangulum II respectively. These are both capable of covering the entire parameter space of both excesses in the event of no radio signal observation by the SKA. Despite the significant $J$-factor and magnetic field uncertainties, these potential limits are at least $2\sigma$ away from the favoured parameter space regions for both excesses. This suggests that very robust constraints can be obtained in future from dwarf spheroidal targets, as is indeed indicated by early observational work in \cite{regis2014b,regis2017}.

\begin{figure}[htbp]
	\centering
	\resizebox{0.49\hsize}{!}{\includegraphics{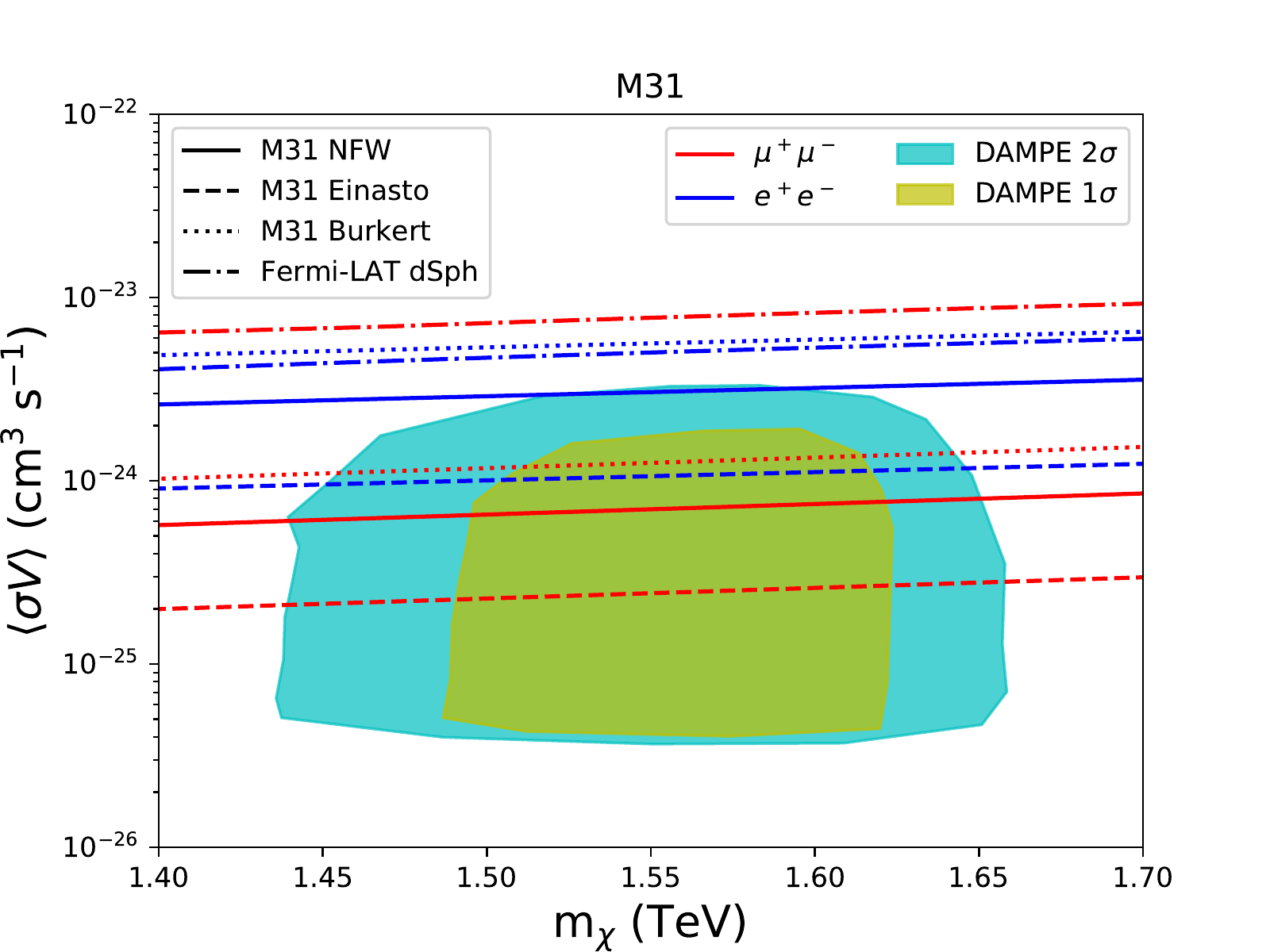}}
	\resizebox{0.49\hsize}{!}{\includegraphics{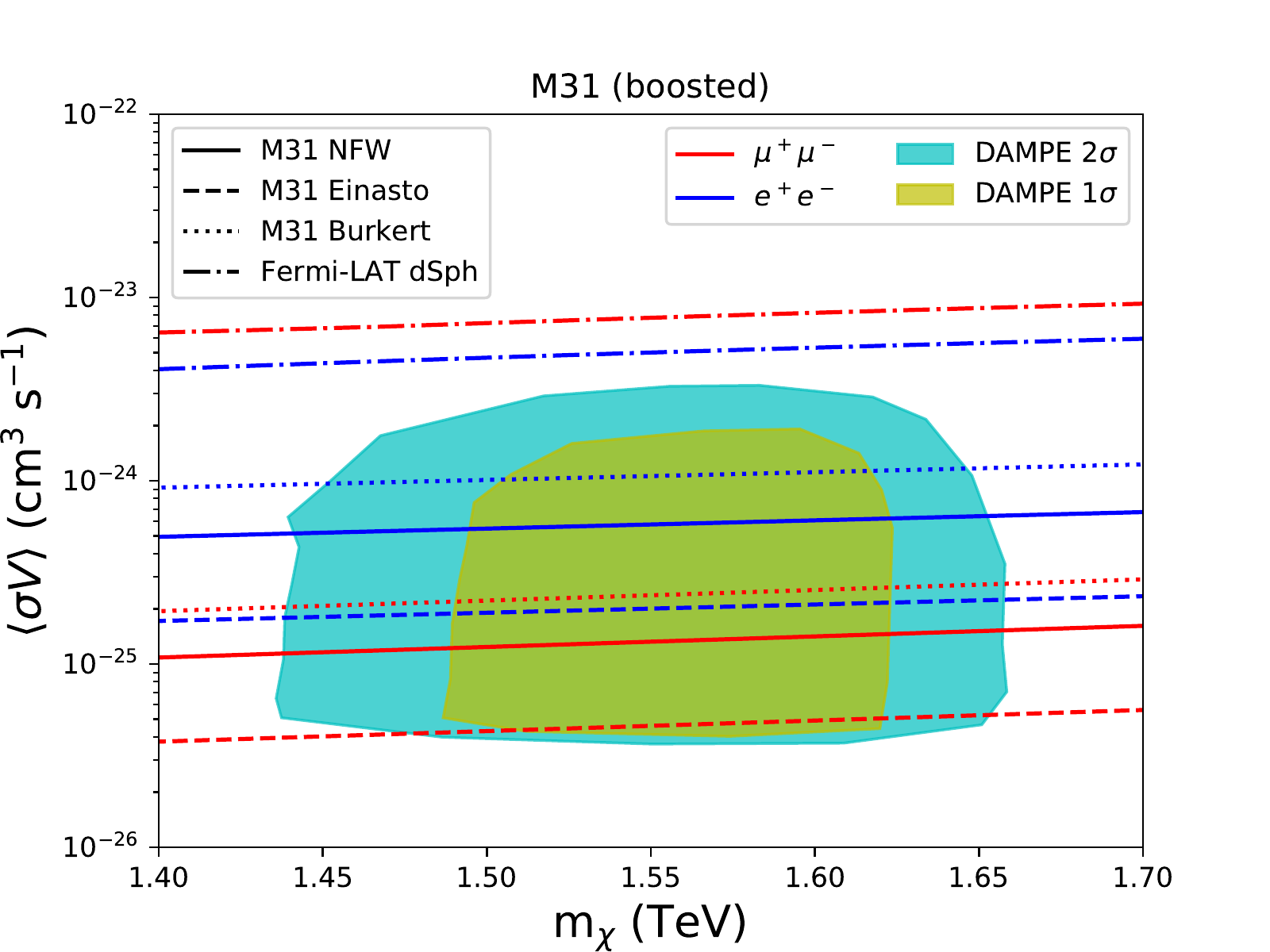}}
	\caption{M31 cross-section upper limits at $2\sigma$ confidence level with 40 kpc ROI. The orange and cyan shaded regions show the $1\sigma$ and $2\sigma$ confidence interval best-fit models for the DAMPE excess from \cite{dampedm1}. The dash-dotted line shows limits from \cite{Fermidwarves2016}. The solid, dotted, and dashed lines show M31 results from this work for NFW, Burkert, and Einasto halos respectively. Note that the muon channel is shown in red and electrons in blue. Left: unboosted. Right: boost factor of $5.28$ used.}
	\label{fig:m31-dampe}
\end{figure}

In Fig.~\ref{fig:m31-dampe} we compare the parameter space region favoured by the DAMPE excess~\cite{dampedm1} with the inclusion of a local DM over-density producing the excess cosmic-rays. This indicates that, due to the large mass of the WIMP, the DAMPE parameter space is challenging to probe. Indeed, as argued in \cite{dampedm1}, it is untouched by Fermi-LAT dwarf galaxy limits and is only strongly impacted by M31 data with NFW (muon channel only) and Einasto halo profile choices, with Burkert halos producing weaker constraints on the parameter space with the muon channel. This is largely due to the sensitivity of the frequency of the peak of DM synchrotron spectrum to WIMP mass~\cite{gsp2015}. In the right panel of Fig.~\ref{fig:m31-dampe} we can see that the principle uncertainty will be the boosting factor, as the use of the value from \cite{chan2019} results in strong constraints for NFW and Einasto halo profiles, with even the Burkert case making in-roads on the parameter space. However, in the pessimistic case where we assume \cite{ficarra1985} has a 5 arcminute ROI we can only minimally constrain the DAMPE parameter space. A similar exercise with the data points from \cite{chan2019} provide constraints around an order of magnitude weaker in each case, often meaning that no limits are put upon the parameter space without a boosting factor (which allows marginal limitation only). 

\begin{figure}[htbp]
	\centering
	\resizebox{0.49\hsize}{!}{\includegraphics{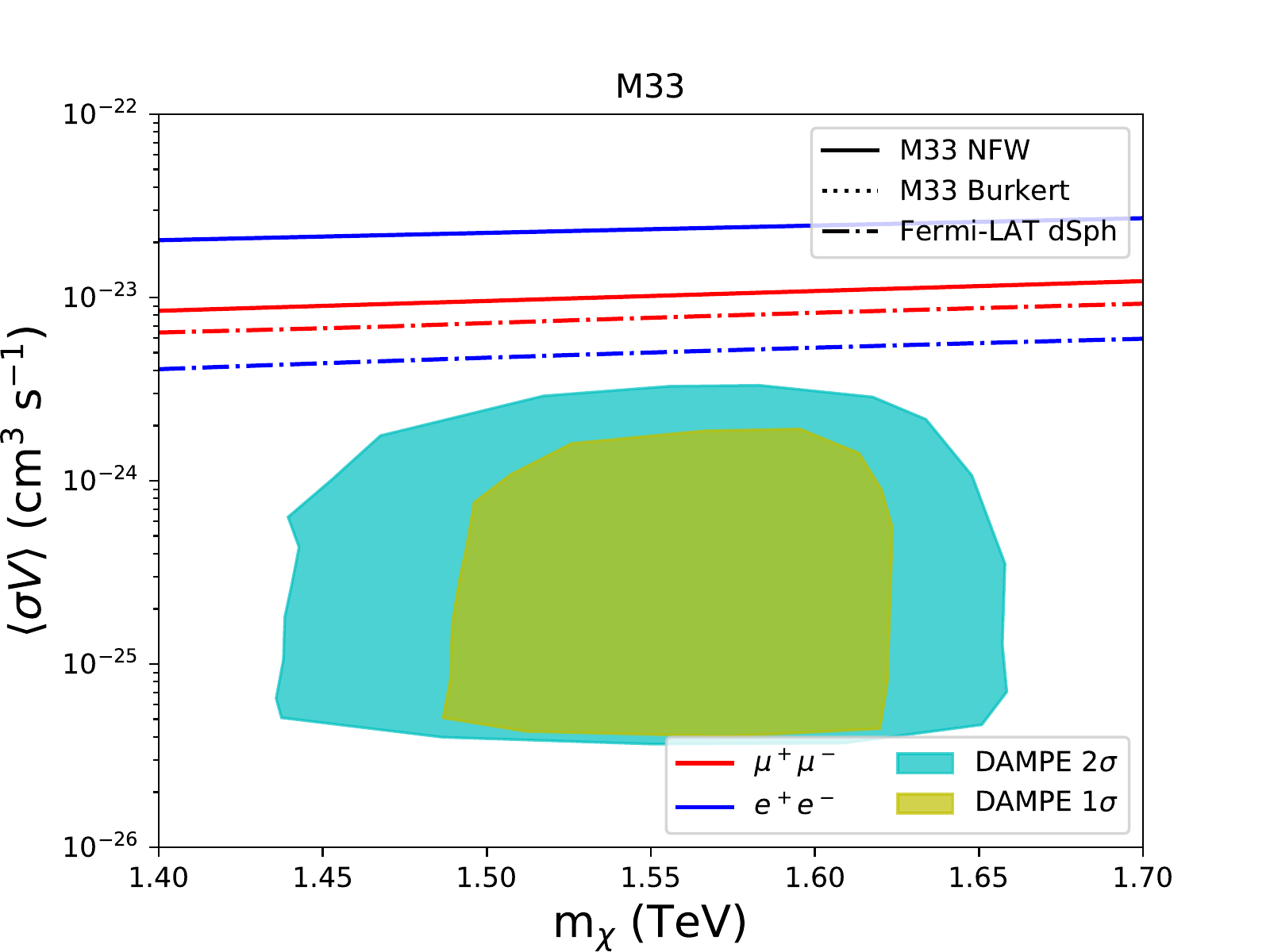}}
	\resizebox{0.49\hsize}{!}{\includegraphics{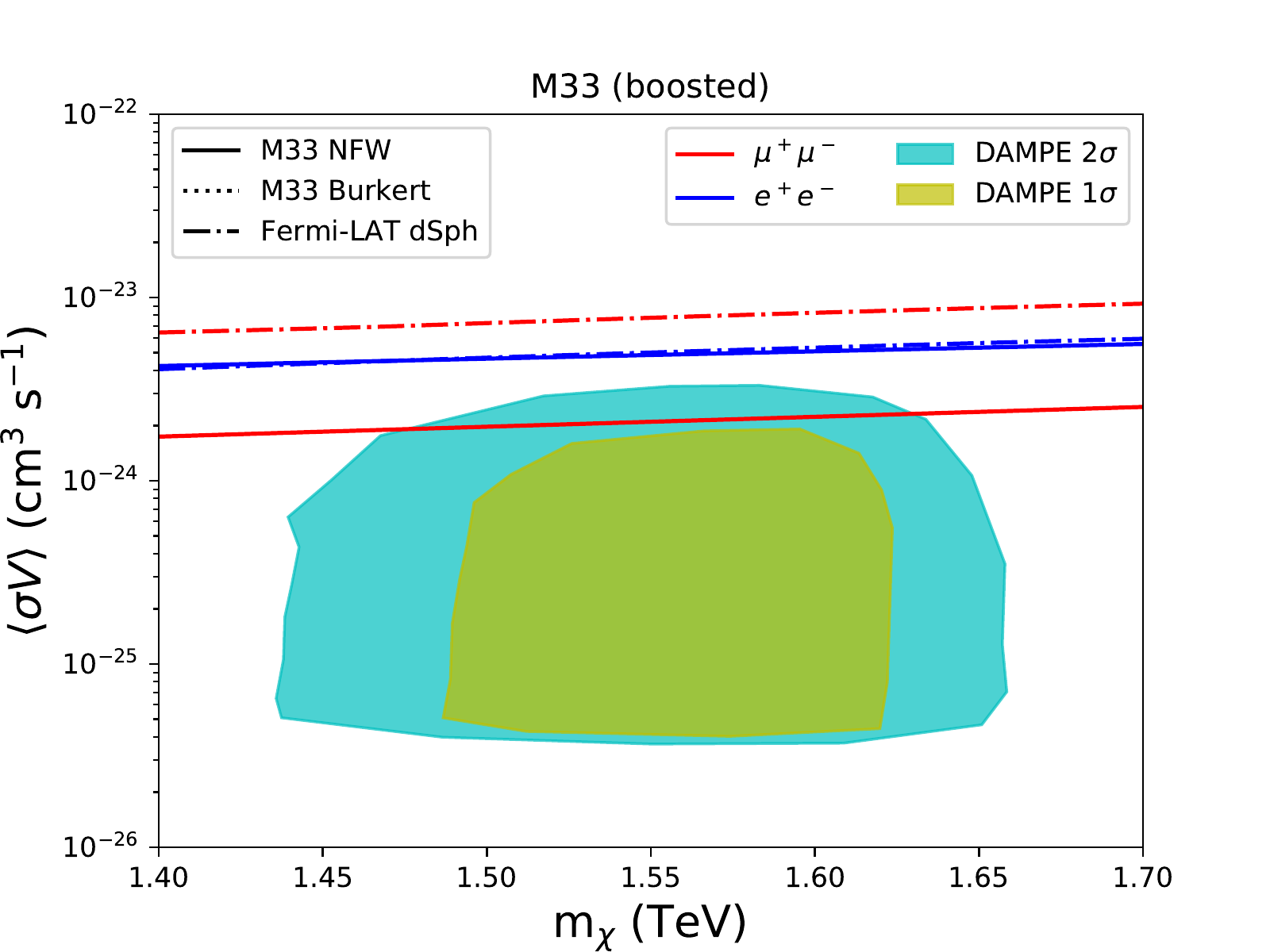}}
	\caption{M33 cross-section upper limits at $2\sigma$ confidence level. The orange and cyan shaded regions show the $1\sigma$ and $2\sigma$ confidence interval best-fit models for the DAMPE excess from \cite{dampedm1}. The dash-dotted line shows limits from \cite{Fermidwarves2016}. The solid and dotted lines show M33 results from this work for NFW and Burkert halos respectively. Note that the muon channel is shown in red and electrons in blue. Left: unboosted. Right: boost factor of $4.86$ used.}
	\label{fig:m33-dampe}
\end{figure}

Figure~\ref{fig:m33-dampe} displays results like those of Fig.~\ref{fig:m31-dampe} but for the case of M33. In the unboosted case even the data from \cite{nvss1998} makes no impact on the parameter space and a boosting factor allows only marginal impact for an NFW halo and the muon channel.

\section{Discussion and Conclusions}
\label{sec:conc}

This work has demonstrated, by studying M31 and M33, that the use of constant magnetic field profiles in calculating DM synchrotron emission should be approached with caution. Interestingly, despite a robust argument in \cite{chan2017} that diffusion should be insignificant in the M33 magnetic environment, we demonstrate that a steep radial profile for the magnetic field strength, and small ROI, make the inclusion of diffusive effects essential. Thus, diffusive effects can only be ignored when both the time-scale argument holds and when the magnetic field profile is sufficiently shallow. Interestingly, our results using data from \cite{chan2017}, for all but electron annihilation channel agree well with those found in \cite{chan2017} despite the large difference in magnetic and diffusive environmental assumptions. The findings presented here reduce the impact of the limits on the DM annihilation cross-section from \cite{chan2017,chan2019} substantially in M31 as well as for electron channel in M33.

However, by expanding the radio data set for M31 and M33, the impact of resulting annihilation limits can be restored largely without recourse to uncertain substructure boosting factors. With the results presented here for M31 from the data enumerated in Section~\ref{sec:data} being very similar to those reported in \cite{chan2019}, with the added advantage of extending down to lower WIMP masses due to the extent of the frequency range covered. In particular, the results presented here for M31 and M33 are shown to be capable of ruling out DM models proposed to account for the Galactic Centre gamma-ray GeV excess and the anti-proton and positron excesses observed by AMS-02~\cite{cholis2019,dimauro2015} (in our pessimistic M31 scenario a boosting factor $\approx 5$ is needed for this). This remains true regardless of the choice of halo density profile in M31, but only for an NFW halo in M33. We note that, for M31, this constitutes a conservative estimate as we use only the extrapolation of magnetic field strength from \cite{ruiz-granados2010} without including an exponential component at the centre of M31. This would be conservative as the flux from DM-based emissions is strongly concentrated in the central regions of the halo by the $\rho^2$ dependence, making the central field value very influential. Additionally, as noted in \cite{ruiz-granados2010}, M31 is likely to have both a central exponential component and the outer flatter one modelled by \cite{ruiz-granados2010}. Despite this, our results agree quite closely with the joint constraints presented for M31 in \cite{egorov2013} (who use an exponential field profile). In particular, the optimistic M31 case matches closely (for an NFW halo profile) with more optimistic central magnetic field strength $B_0 = 50$ $\mu$G results, while the pessimistic case agrees closely with the most pessimistic field estimates from \cite{egorov2013}. We note particularly that our results display the power of low frequency data in constraining DM radio emissions, strongly in agreement with \cite{egorov2013} where the $325$ and $74$ MHz data provide the strongest constraints over most of the studied WIMP mass range (particularly for the $b$ quark channel).

Additionally, we show that the dwarf galaxies Reticulum II and Triangulum II both have the potential to rule out the favoured DM parameter spaces for the Galactic Centre GeV gamma-ray excess as well as for both the AMS-02 positron and anti-proton cases. This despite the inclusion of very significant uncertainties in halo $J$-factors, assuming only 50 hours of observation time, the use of a Burkert density profile, and conservative magnetic field assumptions.

Finally, we employed the M31 and M33 limits to examine the DM models proposed to account for the cosmic-ray excess seen by DAMPE. We find that the favoured parameter space from \cite{dampedm1} cannot be strongly probed via M31 or M33 data without the assumption of a boosting factor. As we are required to reach the thermal relic annihilation cross-section for 1 TeV WIMPs in light lepton annihilation channels and DM radio spectra peak at a frequency sensitive to the WIMP mass. However, for NFW and Einasto density profiles in M31 some impact on the DAMPE DM parameter space is possible, particularly when the WIMP annihilates via muons. When a boosting factor is assumed a large majority of the parameter space can be explored for NFW as well as an Einasto cases for M31. The use of a Burkert density profile produces weaker limits even when a boosting factor is used. Notably, M33 does not strongly impact the parameter space for the DAMPE excess regardless of halo geometry or boosting. This DAMPE DM parameter space can also potentially be explored by the SKA hunting for radio emissions from the nearby DM clump, necessary to the DAMPE excess model from \cite{dampedm1}, as argued in \cite{gSAIP2018}.

\section*{Acknowledgments}
G.B acknowledges support from a National Research Foundation of South Africa Thuthuka grant no. 117969. This research has made use of the NASA/IPAC Extragalactic Database (NED), which is operated by the Jet Propulsion Laboratory, California Institute of Technology, under contract with the National Aeronautics and Space Administration. This work also made use of the WebPlotDigitizer\footnote{\url{http://automeris.io/WebPlotDigitizer/}}.

\bibliographystyle{JHEP}
\bibliography{m31_radio}

\providecommand{\href}[2]{#2}\begingroup\raggedright\begin{thebibliography}{10}

\bibitem{amsdocs-ting2013}
S.~Ting, \emph{{The Alpha Magnetic Spectrometer on the International Space
  Station}},
  \href{https://doi.org/10.1016/j.nuclphysbps.2013.09.028}{\emph{Nucl. Phys.
  Proc. Suppl.} {\bfseries 243-244} (2013) 12}.

\bibitem{ams2-antiprotons-aguilar2016}
{\scshape AMS} collaboration, \emph{Antiproton flux, antiproton-to-proton flux
  ratio, and properties of elementary particle fluxes in primary cosmic rays
  measured with the alpha magnetic spectrometer on the international space
  station}, \href{https://doi.org/10.1103/PhysRevLett.117.091103}{\emph{Phys.
  Rev. Lett.} {\bfseries 117} (2016) 091103}.

\bibitem{pamela-docs}
P.~Picozza et~al., \emph{{PAMELA: A Payload for Antimatter Matter Exploration
  and Light-nuclei Astrophysics}},
  \href{https://doi.org/10.1016/j.astropartphys.2006.12.002}{\emph{Astropart.
  Phys.} {\bfseries 27} (2007) 296}
  [\href{https://arxiv.org/abs/astro-ph/0608697}{{\ttfamily
  astro-ph/0608697}}].

\bibitem{pamela-cr-spectrum2011}
O.~{Adriani}, G.~C. {Barbarino}, G.~A. {Bazilevskaya}, R.~{Bellotti},
  M.~{Boezio}, E.~A. {Bogomolov} et~al., \emph{{PAMELA Measurements of
  Cosmic-Ray Proton and Helium Spectra}},
  \href{https://doi.org/10.1126/science.1199172}{\emph{Science} {\bfseries 332}
  (2011) 69} [\href{https://arxiv.org/abs/1103.4055}{{\ttfamily 1103.4055}}].

\bibitem{dampe-ambrosi2017}
{\scshape DAMPE} collaboration, \emph{{Direct detection of a break in the
  teraelectronvolt cosmic-ray spectrum of electrons and positrons}},
  \href{https://doi.org/10.1038/nature24475}{\emph{Nature} {\bfseries 552}
  (2017) 63} [\href{https://arxiv.org/abs/1711.10981}{{\ttfamily 1711.10981}}].

\bibitem{hess-cr-aharonian2008}
{\scshape H.E.S.S.} collaboration, \emph{{The energy spectrum of cosmic-ray
  electrons at TeV energies}},
  \href{https://doi.org/10.1103/PhysRevLett.101.261104}{\emph{Phys. Rev. Lett.}
  {\bfseries 101} (2008) 261104}
  [\href{https://arxiv.org/abs/0811.3894}{{\ttfamily 0811.3894}}].

\bibitem{fermidocs-atwood2009}
W.~B. {Atwood}, A.~A. {Abdo}, M.~{Ackermann}, W.~{Althouse}, B.~{Anderson},
  M.~{Axelsson} et~al., \emph{{The Large Area Telescope on the Fermi Gamma-Ray
  Space Telescope Mission}},
  \href{https://doi.org/10.1088/0004-637X/697/2/1071}{\emph{ApJ} {\bfseries
  697} (2009) 1071} [\href{https://arxiv.org/abs/0902.1089}{{\ttfamily
  0902.1089}}].

\bibitem{fermi-cr-2017}
{\scshape Fermi-LAT} collaboration, \emph{{Cosmic-ray electron-positron
  spectrum from 7 GeV to 2 TeV with the Fermi Large Area Telescope}},
  \href{https://doi.org/10.1103/PhysRevD.95.082007}{\emph{Phys. Rev.}
  {\bfseries D95} (2017) 082007}
  [\href{https://arxiv.org/abs/1704.07195}{{\ttfamily 1704.07195}}].

\bibitem{dm-cr-bergstrom1999}
L.~Bergstrom, J.~Edsjo and P.~Ullio, \emph{{Cosmic anti-protons as a probe for
  supersymmetric dark matter?}},
  \href{https://doi.org/10.1086/307975}{\emph{Astrophys. J.} {\bfseries 526}
  (1999) 215} [\href{https://arxiv.org/abs/astro-ph/9902012}{{\ttfamily
  astro-ph/9902012}}].

\bibitem{dm-cr-hooper2003}
D.~Hooper, J.~E. Taylor and J.~Silk, \emph{{Can supersymmetry naturally explain
  the positron excess?}},
  \href{https://doi.org/10.1103/PhysRevD.69.103509}{\emph{Phys. Rev.}
  {\bfseries D69} (2004) 103509}
  [\href{https://arxiv.org/abs/hep-ph/0312076}{{\ttfamily hep-ph/0312076}}].

\bibitem{dm-cr-bringmann2006}
T.~Bringmann and P.~Salati, \emph{{The galactic antiproton spectrum at high
  energies: Background expectation vs. exotic contributions}},
  \href{https://doi.org/10.1103/PhysRevD.75.083006}{\emph{Phys. Rev.}
  {\bfseries D75} (2007) 083006}
  [\href{https://arxiv.org/abs/astro-ph/0612514}{{\ttfamily
  astro-ph/0612514}}].

\bibitem{dm-cr-hooper2014}
D.~Hooper, T.~Linden and P.~Mertsch, \emph{{What Does The PAMELA Antiproton
  Spectrum Tell Us About Dark Matter?}},
  \href{https://doi.org/10.1088/1475-7516/2015/03/021}{\emph{JCAP} {\bfseries
  1503} (2015) 021} [\href{https://arxiv.org/abs/1410.1527}{{\ttfamily
  1410.1527}}].

\bibitem{dm-cr-cirelli2014}
M.~Cirelli, D.~Gaggero, G.~Giesen, M.~Taoso and A.~Urbano, \emph{{Antiproton
  constraints on the GeV gamma-ray excess: a comprehensive analysis}},
  \href{https://doi.org/10.1088/1475-7516/2014/12/045}{\emph{JCAP} {\bfseries
  1412} (2014) 045} [\href{https://arxiv.org/abs/1407.2173}{{\ttfamily
  1407.2173}}].

\bibitem{dm-cr-cui2017}
M.-Y. Cui, Q.~Yuan, Y.-L.~S. Tsai and Y.-Z. Fan, \emph{{Possible dark matter
  annihilation signal in the AMS-02 antiproton data}},
  \href{https://doi.org/10.1103/PhysRevLett.118.191101}{\emph{Phys. Rev. Lett.}
  {\bfseries 118} (2017) 191101}
  [\href{https://arxiv.org/abs/1610.03840}{{\ttfamily 1610.03840}}].

\bibitem{dm-cr-cuoco2017a}
A.~Cuoco, M.~Kr{\"{a}}mer and M.~Korsmeier, \emph{{Novel Dark Matter
  Constraints from Antiprotons in Light of AMS-02}},
  \href{https://doi.org/10.1103/PhysRevLett.118.191102}{\emph{Phys. Rev. Lett.}
  {\bfseries 118} (2017) 191102}
  [\href{https://arxiv.org/abs/1610.03071}{{\ttfamily 1610.03071}}].

\bibitem{dm-cr-cuoco2017b}
A.~Cuoco, J.~Heisig, M.~Korsmeier and M.~Kr{\"{a}}mer, \emph{Probing dark
  matter annihilation in the galaxy with antiprotons and gamma rays},
  \href{https://doi.org/10.1088/1475-7516/2017/10/053}{\emph{JCAP} {\bfseries
  1710} (2017) 053} [\href{https://arxiv.org/abs/1704.08258}{{\ttfamily
  1704.08258}}].

\bibitem{cholis2019}
I.~Cholis, T.~Linden and D.~Hooper, \emph{{A Robust Excess in the Cosmic-Ray
  Antiproton Spectrum: Implications for Annihilating Dark Matter}},
  \href{https://arxiv.org/abs/1903.02549}{{\ttfamily 1903.02549}}.

\bibitem{dimauro2015}
M.~Di~Mauro, F.~Donato, N.~Fornengo and A.~Vittino, \emph{{Dark matter vs.
  astrophysics in the interpretation of AMS-02 electron and positron data}},
  \href{https://doi.org/10.1088/1475-7516/2016/05/031}{\emph{JCAP} {\bfseries
  1605} (2016) 031} [\href{https://arxiv.org/abs/1507.07001}{{\ttfamily
  1507.07001}}].

\bibitem{carena2019}
M.~Carena, J.~Osborne, N.~R. Shah and C.~E.~M. Wagner, \emph{{The Return of the
  WIMP: Missing Energy Signals and the Galactic Center Excess}},
  \href{https://arxiv.org/abs/1905.03768}{{\ttfamily 1905.03768}}.

\bibitem{calore2015}
F.~Calore, I.~Cholis and C.~Weniger, \emph{{Background Model Systematics for
  the Fermi GeV Excess}},
  \href{https://doi.org/10.1088/1475-7516/2015/03/038}{\emph{JCAP} {\bfseries
  1503} (2015) 038} [\href{https://arxiv.org/abs/1409.0042}{{\ttfamily
  1409.0042}}].

\bibitem{oleary2015}
R.~M. O'Leary, M.~D. Kistler, M.~Kerr and J.~Dexter, \emph{{Young Pulsars and
  the Galactic Center GeV Gamma-ray Excess}},
  \href{https://arxiv.org/abs/1504.02477}{{\ttfamily 1504.02477}}.

\bibitem{bartels2015}
R.~Bartels, S.~Krishnamurthy and C.~Weniger, \emph{{Strong support for the
  millisecond pulsar origin of the Galactic center GeV excess}},
  \href{https://doi.org/10.1103/PhysRevLett.116.051102}{\emph{Phys. Rev. Lett.}
  {\bfseries 116} (2016) 051102}
  [\href{https://arxiv.org/abs/1506.05104}{{\ttfamily 1506.05104}}].

\bibitem{lee2015}
S.~K. Lee, M.~Lisanti, B.~R. Safdi, T.~R. Slatyer and W.~Xue, \emph{{Evidence
  for Unresolved $\gamma$-Ray Point Sources in the Inner Galaxy}},
  \href{https://doi.org/10.1103/PhysRevLett.116.051103}{\emph{Phys. Rev. Lett.}
  {\bfseries 116} (2016) 051103}
  [\href{https://arxiv.org/abs/1506.05124}{{\ttfamily 1506.05124}}].

\bibitem{brandt2015}
T.~D. Brandt and B.~Kocsis, \emph{{Disrupted Globular Clusters Can Explain the
  Galactic Center Gamma Ray Excess}},
  \href{https://doi.org/10.1088/0004-637X/812/1/15}{\emph{Astrophys. J.}
  {\bfseries 812} (2015) 15}
  [\href{https://arxiv.org/abs/1507.05616}{{\ttfamily 1507.05616}}].

\bibitem{gs2016}
G.~Beck and S.~Colafrancesco, \emph{{A Multi-frequency analysis of dark matter
  annihilation interpretations of recent anti-particle and $\gamma$-ray
  excesses in cosmic structures}},
  \href{https://doi.org/10.1088/1475-7516/2016/05/013}{\emph{JCAP} {\bfseries
  1605} (2016) 013} [\href{https://arxiv.org/abs/1508.01386}{{\ttfamily
  1508.01386}}].

\bibitem{dampedm1}
Y.-Z. Fan, W.-C. Huang, M.~Spinrath, Y.-L.~S. Tsai and Q.~Yuan, \emph{{A model
  explaining neutrino masses and the DAMPE cosmic ray electron excess}},
  \href{https://doi.org/10.1016/j.physletb.2018.03.066}{\emph{Phys. Lett.}
  {\bfseries B781} (2018) 83}
  [\href{https://arxiv.org/abs/1711.10995}{{\ttfamily 1711.10995}}].

\bibitem{dampedm2}
Q.~Yuan et~al., \emph{{Interpretations of the DAMPE electron data}},
  \href{https://arxiv.org/abs/1711.10989}{{\ttfamily 1711.10989}}.

\bibitem{dampeucmh}
F.~Yang, M.~Su and Y.~Zhao, \emph{{Dark Matter Annihilation from Nearby
  Ultra-compact Micro Halos to Explain the Tentative Excess at ~1.4 TeV in
  DAMPE data}},  \href{https://arxiv.org/abs/1712.01724}{{\ttfamily
  1712.01724}}.

\bibitem{gsp2015}
S.~Colafrancesco, P.~Marchegiani and G.~Beck, \emph{{Evolution of Dark Matter
  Halos and their Radio Emissions}},
  \href{https://doi.org/10.1088/1475-7516/2015/02/032}{\emph{JCAP} {\bfseries
  1502} (2015) 032} [\href{https://arxiv.org/abs/1409.4691}{{\ttfamily
  1409.4691}}].

\bibitem{Colafranceso2015}
S.~Colafrancesco, M.~Regis, P.~Marchegiani, G.~Beck, R.~Beck, H.~Zechlin
  et~al., \emph{{Probing the nature of Dark Matter with the SKA}},
  \href{https://doi.org/10.22323/1.215.0100}{\emph{PoS} {\bfseries AASKA14}
  (2015) 100} [\href{https://arxiv.org/abs/1502.03738}{{\ttfamily
  1502.03738}}].

\bibitem{Colafrancesco2007}
S.~Colafrancesco, S.~Profumo and P.~Ullio, \emph{Detecting dark matter wimps in
  the draco dwarf: a multi-wavelength perspective}, {\emph{Phys. Rev. D}
  {\bfseries 75} (2007) 023513}.

\bibitem{chan2017}
M.~H. Chan, \emph{{Constraining annihilating dark matter by radio data of
  M33}}, \href{https://doi.org/10.1103/PhysRevD.96.043009}{\emph{Phys. Rev.}
  {\bfseries D96} (2017) 043009}
  [\href{https://arxiv.org/abs/1708.01370}{{\ttfamily 1708.01370}}].

\bibitem{chan2019}
M.~H. Chan, L.~Cui, J.~Liu and C.~S. Leung, \emph{{Ruling out $\sim 100-300$
  GeV thermal relic annihilating dark matter by radio observation of the
  Andromeda galaxy}},
  \href{https://doi.org/10.3847/1538-4357/aafe0b}{\emph{ApJ} {\bfseries 872}
  (2019) 177} [\href{https://arxiv.org/abs/1901.04638}{{\ttfamily
  1901.04638}}].

\bibitem{regis2014b}
M.~Regis, L.~Richter, S.~Colafrancesco, S.~Profumo, W.~J.~G. de Blok and
  M.~Massardi, \emph{Local group dsph radio survey with atca - ii. non-thermal
  diffuse emission}, \href{https://doi.org/10.1093/mnras/stv127}{\emph{Monthly
  Notices of the Royal Astronomical Society} {\bfseries 448} (2015) 3747}.

\bibitem{regis2017}
M.~Regis, L.~Richter and S.~Colafrancesco, \emph{Dark matter in the reticulum
  ii dsph: a radio search}, {\emph{Journal of Cosmology and Astroparticle
  Physics} {\bfseries 2017} (2017) 025}.

\bibitem{gb2019a}
G.~Beck, \emph{Radio-frequency searches for dark matter in dwarf galaxies},
  \href{https://doi.org/10.3390/galaxies7010016}{\emph{Galaxies} {\bfseries 7}
  (2019) }.

\bibitem{abazajian2016}
K.~N. Abazajian and R.~E. Keeley, \emph{Bright gamma-ray galactic center excess
  and dark dwarfs: Strong tension for dark matter annihilation despite milky
  way halo profile and diffuse emission uncertainties},
  \href{https://doi.org/10.1103/PhysRevD.93.083514}{\emph{Phys. Rev. D}
  {\bfseries 93} (2016) 083514}.

\bibitem{ficarra1985}
A.~{Ficarra}, G.~{Grueff} and G.~{Tomassetti}, \emph{{A new Bologna sky survey
  at 408 MHz}}, {\emph{Astronomy and Astrophysics Supplement Series} {\bfseries
  59} (1985) 255}.

\bibitem{ned_m31_rice1988}
W.~{Rice}, C.~J. {Lonsdale}, B.~T. {Soifer}, G.~{Neugebauer}, E.~L. {Kopan},
  L.~A. {Lloyd} et~al., \emph{{A catalog of IRAS observations of large optical
  galaxies}}, \href{https://doi.org/10.1086/191283}{\emph{ApJS} {\bfseries 68}
  (1988) 91}.

\bibitem{ned_m31_condon2002}
J.~J. {Condon}, W.~D. {Cotton} and J.~J. {Broderick}, \emph{{Radio Sources and
  Star Formation in the Local Universe}},
  \href{https://doi.org/10.1086/341650}{\emph{The Astronomical Journal}
  {\bfseries 124} (2002) 675}.

\bibitem{ned_m31_jarrett2003}
T.~H. {Jarrett}, T.~{Chester}, R.~{Cutri}, S.~E. {Schneider} and J.~P.
  {Huchra}, \emph{{The 2MASS Large Galaxy Atlas}},
  \href{https://doi.org/10.1086/345794}{\emph{The Astronomical Journal}
  {\bfseries 125} (2003) 525}.

\bibitem{ned_m31_still2009}
J.~M. {Stil}, M.~{Krause}, R.~{Beck} and A.~R. {Taylor}, \emph{{The Integrated
  Polarization of Spiral Galaxy Disks}},
  \href{https://doi.org/10.1088/0004-637X/693/2/1392}{\emph{ApJ} {\bfseries
  693} (2009) 1392} [\href{https://arxiv.org/abs/0810.2303}{{\ttfamily
  0810.2303}}].

\bibitem{nvss1998}
J.~J. {Condon}, W.~D. {Cotton}, E.~W. {Greisen}, Q.~F. {Yin}, R.~A. {Perley},
  G.~B. {Taylor} et~al., \emph{{The NRAO VLA Sky Survey}},
  \href{https://doi.org/10.1086/300337}{\emph{{The Astronomical Journal}}
  {\bfseries 115} (1998) 1693}.

\bibitem{steigman2012}
G.~Steigman, B.~Dasgupta and J.~F. Beacom, \emph{{Precise Relic WIMP Abundance
  and its Impact on Searches for Dark Matter Annihilation}},
  \href{https://doi.org/10.1103/PhysRevD.86.023506}{\emph{Phys. Rev.}
  {\bfseries D86} (2012) 023506}
  [\href{https://arxiv.org/abs/1204.3622}{{\ttfamily 1204.3622}}].

\bibitem{egorov2013}
A.~E. {Egorov} and E.~{Pierpaoli}, \emph{{Constraints on dark matter
  annihilation by radio observations of M31}},
  \href{https://doi.org/10.1103/PhysRevD.88.023504}{\emph{Phys. Rev. D}
  {\bfseries 88} (2013) 023504}
  [\href{https://arxiv.org/abs/1304.0517}{{\ttfamily 1304.0517}}].

\bibitem{ppdmcb1}
M.~Cirelli et~al., \emph{Pppc 4 dm id: A poor particle physicist cookbook for
  dark matter indirect detection}, {\emph{JCAP} {\bfseries 1103} (2011) 051}.

\bibitem{ppdmcb2}
P.~Ciafaloni et~al., \emph{Weak corrections are relevant for dark matter
  indirect detection}, {\emph{JCAP} {\bfseries 1103} (2011) 019}.

\bibitem{longair1994}
M.~S. Longair, \emph{High Energy Astrophysics}. Cambridge University Press,
  1994.

\bibitem{rybicki1986}
G.~B. {Rybicki} and A.~P. {Lightman}, \emph{{Radiative Processes in
  Astrophysics}}. Wiley, June, 1986.

\bibitem{Colafrancesco2006}
S.~Colafrancesco, S.~Profumo and P.~Ullio, \emph{Multi-frequency analysis of
  neutralino dark matter annihilations in the coma cluster}, {\emph{A\&A}
  {\bfseries 455} (2006) 21}.

\bibitem{baltz1999}
E.~A. Baltz and J.~Edsj\"o, \emph{Positron propagation and fluxes from
  neutralino annihilation in the halo},
  \href{https://doi.org/10.1103/PhysRevD.59.023511}{\emph{Phys. Rev. D}
  {\bfseries 59} (1998) 023511}.

\bibitem{baltz2004}
E.~A. Baltz and L.~Wai, \emph{Diffuse inverse compton and synchrotron emission
  from dark matter annihilations in galactic satellites},
  \href{https://doi.org/10.1103/PhysRevD.70.023512}{\emph{Phys. Rev. D}
  {\bfseries 70} (2004) 023512}.

\bibitem{Colafrancesco1998}
S.~Colafrancesco and S.~Blasi, \emph{Clusters of galaxies and the diffuse gamma
  ray background}, {\emph{Astropart. Phys.} {\bfseries 9} (1998) 227}.

\bibitem{tamm2012}
A.~{Tamm}, E.~{Tempel}, P.~{Tenjes}, O.~{Tihhonova} and T.~{Tuvikene},
  \emph{{Stellar mass map and dark matter distribution in M 31}},
  \href{https://doi.org/10.1051/0004-6361/201220065}{\emph{A\&A} {\bfseries
  546} (2012) A4} [\href{https://arxiv.org/abs/1208.5712}{{\ttfamily
  1208.5712}}].

\bibitem{nfw1996}
J.~F. Navarro, C.~S. Frenk and S.~D.~M. White, \emph{{The Structure of cold
  dark matter halos}}, \href{https://doi.org/10.1086/177173}{\emph{Astrophys.
  J.} {\bfseries 462} (1996) 563}
  [\href{https://arxiv.org/abs/astro-ph/9508025}{{\ttfamily
  astro-ph/9508025}}].

\bibitem{burkert1995}
A.~Burkert, \emph{{The Structure of dark matter halos in dwarf galaxies}},
  \href{https://doi.org/10.1086/309560}{\emph{IAU Symp.} {\bfseries 171} (1996)
  175} [\href{https://arxiv.org/abs/astro-ph/9504041}{{\ttfamily
  astro-ph/9504041}}].

\bibitem{einasto1968}
J.~Einasto, \emph{On galactic descriptive functions}, {\emph{Publications of
  the Tartuskoj Astrofizica Observatory} {\bfseries 36} (1968) 414}.

\bibitem{ruiz-granados2010}
B.~Ruiz-Granados, J.~A. Rubi{\~{n}}o-Mart{\'{\i}}n, E.~Florido and E.~Battaner,
  \emph{Magnetic fields and the outer rotation curve of m31},
  \href{https://doi.org/10.1088/2041-8205/723/1/l44}{\emph{ApJ} {\bfseries 723}
  (2010) L44}.

\bibitem{beckm31}
R.~Beck, \emph{The magnetic field in m31}, {\emph{A\&A} {\bfseries 106} (1982)
  121}.

\bibitem{fune2016}
E.~Lopez~Fune, P.~Salucci and E.~Corbelli, \emph{{Radial dependence of the dark
  matter distribution in M33}},
  \href{https://doi.org/10.1093/mnras/stx429}{\emph{Mon. Not. Roy. Astron.
  Soc.} {\bfseries 468} (2017) 147}
  [\href{https://arxiv.org/abs/1611.01409}{{\ttfamily 1611.01409}}].

\bibitem{beck2015}
R.~{Beck}, \emph{{Magnetic fields in spiral galaxies}},
  \href{https://doi.org/10.1007/s00159-015-0084-4}{\emph{The Astronomy and
  Astrophysics Review} {\bfseries 24} (2015) 4}
  [\href{https://arxiv.org/abs/1509.04522}{{\ttfamily 1509.04522}}].

\bibitem{berkhuijsen2013}
E.~M. Berkhuijsen, R.~Beck and F.~S. Tabatabaei, \emph{{How cosmic-ray electron
  propagation affects radio-far-infrared correlations in M31 and M33}},
  \href{https://doi.org/10.1093/mnras/stt1400}{\emph{Mon. Not. Roy. Astron.
  Soc.} {\bfseries 435} (2013) 1598}
  [\href{https://arxiv.org/abs/1307.7991}{{\ttfamily 1307.7991}}].

\bibitem{regan1994}
M.~W. {Regan} and S.~N. {Vogel}, \emph{{The near-infrared structure of M33}},
  \href{https://doi.org/10.1086/174755}{\emph{ApJ} {\bfseries 434} (1994) 536}.

\bibitem{walker2009}
M.~G. Walker, M.~Mateo, E.~W. Olszewski, J.~P. narrubia, N.~W. Evans and
  G.~Gilmore, \emph{A universal mass profile for dwarf spheroidal galaxies?},
  {\emph{ApJ} {\bfseries 704} (2009) 1274}.

\bibitem{adams2014}
J.~J. Adams et~al., \emph{Dwarf galaxy dark matter density profiles inferred
  from stellar and gas kinematics}, {\emph{ApJ} {\bfseries 789} (2014) 63}.

\bibitem{bonnivard2015}
V.~Bonnivard, C.~Combet, D.~Maurin, A.~Geringer-Sameth, S.~M. Koushiappas,
  M.~G. Walker et~al., \emph{{Dark matter annihilation and decay profiles for
  the Reticulum II dwarf spheroidal galaxy}},
  \href{https://doi.org/10.1088/2041-8205/808/2/L36}{\emph{Astrophys. J.}
  {\bfseries 808} (2015) L36}
  [\href{https://arxiv.org/abs/1504.03309}{{\ttfamily 1504.03309}}].

\bibitem{genina2016}
A.~Genina and M.~Fairbairn, \emph{{The potential of the dwarf galaxy Triangulum
  II for dark matter indirect detection}},
  \href{https://doi.org/10.1093/mnras/stw2284}{\emph{Monthly Notices of the
  Royal Astronomical Society} {\bfseries 463} (2016) 3630}
  [\href{https://arxiv.org/abs/http://oup.prod.sis.lan/mnras/article-pdf/463/4/3630/18516441/stw2284.pdf}{{\ttfamily
  http://oup.prod.sis.lan/mnras/article-pdf/463/4/3630/18516441/stw2284.pdf}}].

\bibitem{bechtol2015}
K.~{Bechtol}, A.~{Drlica-Wagner}, E.~{Balbinot}, A.~{Pieres}, J.~D. {Simon},
  B.~{Yanny} et~al., \emph{{Eight New Milky Way Companions Discovered in
  First-year Dark Energy Survey Data}},
  \href{https://doi.org/10.1088/0004-637X/807/1/50}{\emph{ApJ} {\bfseries 807}
  (2015) 50} [\href{https://arxiv.org/abs/1503.02584}{{\ttfamily 1503.02584}}].

\bibitem{koposov2015}
S.~E. Koposov, V.~Belokurov, G.~Torrealba and N.~W. Evans, \emph{{Beasts of the
  Southern Wild: Discovery of nine Ultra Faint satellites in the vicinity of
  the Magellanic Clouds}},
  \href{https://doi.org/10.1088/0004-637X/805/2/130}{\emph{ApJ} {\bfseries 805}
  (2015) 130} [\href{https://arxiv.org/abs/1503.02079}{{\ttfamily
  1503.02079}}].

\bibitem{laevens2015}
B.~P.~M. Laevens, N.~F. Martin, R.~A. Ibata, H.-W. Rix, E.~J. Bernard, E.~F.
  Bell et~al., \emph{A new faint milky way satellite discovered in the
  pan-starrs1 3$\uppi$ survey},
  \href{https://doi.org/10.1088/2041-8205/802/2/l18}{\emph{ApJ} {\bfseries 802}
  (2015) L18}.

\bibitem{strigari2007}
L.~E. {Strigari}, J.~S. {Bullock}, M.~{Kaplinghat}, J.~{Diemand}, M.~{Kuhlen}
  and P.~{Madau}, \emph{{Redefining the Missing Satellites Problem}},
  \href{https://doi.org/10.1086/521914}{\emph{ApJ} {\bfseries 669} (2007) 676}
  [\href{https://arxiv.org/abs/0704.1817}{{\ttfamily 0704.1817}}].

\bibitem{moline2017}
{\'A}.~{Molin{\'e}}, M.~A. {S{\'a}nchez-Conde}, S.~{Palomares-Ruiz} and
  F.~{Prada}, \emph{{Characterization of subhalo structural properties and
  implications for dark matter annihilation signals}},
  \href{https://doi.org/10.1093/mnras/stx026}{\emph{Monthly Notices of the
  Royal Astronomical Society} {\bfseries 466} (2017) 4974}
  [\href{https://arxiv.org/abs/1603.04057}{{\ttfamily 1603.04057}}].

\bibitem{ska2012}
P.~Dewdney, W.~Turner, R.~Millenaar, R.~McCool, J.~Lazio and T.~Cornwell,
  \emph{Ska baseline design document:
  {\url{http://www.skatelescope.org/wp-content/uploads/2012/07/SKA-TEL-SKO-DD-001-1_BaselineDesign1.pdf}}},
  .

\bibitem{Fermidwarves2015}
A.~Drlica-Wagner et~al., \emph{Search for gamma-ray emission from des dwarf
  spheroidal galaxy candidates with fermi-lat data}, {\emph{ApJ} {\bfseries
  809} (2015) L4}.

\bibitem{Fermidwarves2016}
{\scshape Fermi-LAT, DES} collaboration, \emph{{Searching for Dark Matter
  Annihilation in Recently Discovered Milky Way Satellites with Fermi-LAT}},
  \href{https://doi.org/10.3847/1538-4357/834/2/110}{\emph{ApJ} {\bfseries 834}
  (2017) 110} [\href{https://arxiv.org/abs/1611.03184}{{\ttfamily
  1611.03184}}].

\bibitem{siffert2011}
B.~B. {Siffert}, A.~{Limone}, E.~{Borriello}, G.~{Longo} and G.~{Miele},
  \emph{{Radio emission from dark matter annihilation in the Large Magellanic
  Cloud}},
  \href{https://doi.org/10.1111/j.1365-2966.2010.17613.x}{\emph{Monthly Notices
  of the Royal Astronomical Society} {\bfseries 410} (2011) 2463}
  [\href{https://arxiv.org/abs/1006.5325}{{\ttfamily 1006.5325}}].

\bibitem{sofue2015}
Y.~Sofue, \emph{{Dark halos of M31 and the Milky Way}},
  \href{https://doi.org/10.1093/pasj/psv042}{\emph{Publications of the
  Astronomical Society of Japan} {\bfseries 67} (2015) }
  [\href{https://arxiv.org/abs/http://oup.prod.sis.lan/pasj/article-pdf/67/4/75/5166610/psv042.pdf}{{\ttfamily
  http://oup.prod.sis.lan/pasj/article-pdf/67/4/75/5166610/psv042.pdf}}].

\bibitem{gSAIP2018}
G.~Beck and S.~Colafrancesco, \emph{{Dark matter gets DAMPE}},  in \emph{{63rd
  Annual Conference of the South African Institute of Physics (SAIP),
  Bloemfontein, South Africa}}, 2018,
  \href{https://arxiv.org/abs/1810.07176}{{\ttfamily 1810.07176}}.

\end{thebibliography}\endgroup

\end{document}